\documentclass[twocolumn,aps,showpacs,floatfix,nofootinbib]{revtex4-1}
\usepackage{amssymb,amsmath}
\usepackage{hyperref,graphicx}
\usepackage{dcolumn}
\usepackage{color}

\usepackage{latexsym}
\usepackage{bm}
\usepackage{bbm}
\usepackage{psfrag}

\newcommand{\bs}{\boldsymbol}
\newcommand{\bk}{{\boldsymbol k}}
\newcommand{\br}{{\boldsymbol r}}
\newcommand{\bR}{{\boldsymbol R}}

\newcommand{\bv}{{\boldsymbol v}}

\newcommand{\bB}{{\boldsymbol B}}

\newcommand{\bE}{{\boldsymbol E}}

\newcommand{\bx}{{\boldsymbol x}}
\newcommand{\by}{{\boldsymbol y}}
\newcommand{\bz}{{\boldsymbol z}}

\newcommand{\bnabla}{{\boldsymbol\nabla}}

\newcommand{\calTn}[1]{\mathcal{T}^{(#1)}}

\newcommand{\ket}[1]{\left| #1 \right\rangle}
\newcommand{\bra}[1]{\left\langle #1 \right|}

\newcommand{\mv}[1]{\left\langle #1 \right\rangle}

\begin{document}
\title{Multiple scattering of light in cold atomic clouds with a
magnetic field}
\author{Olivier Sigwarth$^{1}$, Guillaume Labeyrie$^{2}$, Dominique Delande$^{3}$, Christian Miniatura$^{2,4,5,6}$
}                     % Do not remove
\affiliation{
\mbox{$^1$ Laboratoire de Photonique, LPO Jean Mermoz, 53 rue du Dr Hurst, 68300 Saint-Louis, France} \\
\mbox{$^2$ Institut Non Lin\'{e}aire de Nice, UMR 6618, UNS, CNRS; 1361 route des Lucioles, 06560 Valbonne, France}\\
\mbox{$^3$ Laboratoire Kastler Brossel, Ecole Normale Sup\'{e}rieure, CNRS, UPMC; 4 Place Jussieu, 75005 Paris, France} \\
\mbox{$^4$ Centre for Quantum Technologies, National University of Singapore, 3 Science Drive 2, Singapore 117543, Singapore} \\
\mbox{$^5$ Department of Physics, National University of Singapore, 2 Science Drive 3, Singapore 117542, Singapore} \\
\mbox{$^6$ Institute of Advanced Studies, NEC, Nanyang Technological University, 60 Nanyang View
Singapore 639673, Singapore} \\
}  

%\institute{Laboratoire de Photonique, LPO Jean Mermoz, 53 rue du Dr Hurst, 68300 Saint-Louis, France \and Institut Non Lin\'eaire de Nice, UNS, CNRS; 1361 route des Lucioles, F-06560 Valbonne, France \and Laboratoire Kastler Brossel, CNRS, UPMC, ENS; 4 Place Jussieu, F-75005 Paris, France \and Centre for Quantum Technologies, National University of Singapore, 3 Science Drive 2, Singapore 117543, Singapore \and Department of Physics, National University of Singapore, 2 Science Drive 3, Singapore 117542, Singapore }
\begin{abstract}
Starting from a microscopic theory for atomic scatterers, we describe the scattering of light by a single atom and study the coherent propagation of light in a cold atomic cloud in the presence of a magnetic field $\bs{B}$ in the mesoscopic regime. Non-pertubative expressions in $B$ are given for the magneto-optical effects and optical anisotropy. We then consider the multiple scattering regime and address the fate of the coherent backscattering (CBS) effect. We show that, for atoms with nonzero spin in their ground state, the CBS interference contrast can be increased compared to its value when $B=0$, a result at variance with classical samples. We validate our theoretical results by a quantitative comparison with experimental data.
\end{abstract}
\pacs{
      {42.25.Dd}, %{Wave propagation in random media}
      {33.57.+c}, %{Magneto-optical and electro-optical spectra and effects}
      {05.60.Gg} %{Quantum transport}
     } % end of PACS codes

\date{\today}

\maketitle
\section{Physical context}
\label{intro}
Multiple scattering of waves is an important topic involved in many branches of physics, from atomic physics to astronomy via condensed matter physics \cite{Chandra,AkkMon}. One of its most fascinating aspects occurs when interference effects persist in the presence of disorder. These interference effects manifest themselves through deviations from the usual diffusive behaviour obtained at long times (weak localization). Under suitable conditions they can even completely inhibit transport (strong localization) reaching an insulating regime. During the last two decades, both the weak and strong localization regimes have been extensively studied with systems ranging from electronic waves to ultracold atoms \cite{Houches,Tourin,Bart_seism,Chabanov,Maret_corr,CBS,Aegerter,Hu,Billy,Chabe}. 

Coherent multiple scattering of light in cold atomic clouds has been experimentally and theoretically studied since ten years in the context of the coherent backscattering (CBS) effect \cite{StructInt}. Compared to classical scattering media, such as semi-conducting powders, cold atoms confined in a magneto-optical trap (MOT) constitute quite a unique sample of identical strongly-resonant point-like scatterers \cite{DiffAt}. A key feature of the atomic scatterers is the presence of quantum internal degrees of freedom when their ground state is degenerate i.e. possesses a non-zero angular momentum \cite{AtVsCl}. These internal degrees of freedom are coupled to the polarization of light waves during single scattering events. As they are traced out when one observes the interference pattern of light, light experiences decoherence during its diffusive propagation in the atomic cloud. This leads to a reduction of the interference between scattering amplitudes as exemplified by the CBS contrast \cite{josab}. The saturation of the atomic transition \cite{Sat} and the finite temperature of the atomic gas \cite{DynBD} are two additional mechanisms reducing the coherence of light. To restore the contrast of the interferences, it has been suggested to use a polarized atomic gas \cite{PolarAt}. Unfortunately, this method is efficient only in optically thin clouds. Another possibility is to break the degeneracy of the atomic ground state by applying an external magnetic field $\bB$, which is one of the topics we address here.

In classical scattering media, the presence of an external magnetic field is known to modify the interference pattern and to reduce the CBS interference contrast \cite{CBSBexp}. The mechanism at work here is the modification of the polarization of light during its coherent propagation due to the Faraday effect \cite{CBSBthMartinez,CBSBthBart}. As this modification is not the same along a scattering path and its reversed partner, a dephasing process takes place between the scattering amplitudes associated to these two paths, scrambling in turn their interference. Although the magneto-optical effects are very strong in cold atomic gases \cite{VoigtCold}, they actually have a negligible impact on the shape of the interference pattern itself \cite{Hanle}. More surprisingly, a full restoration of the interference contrast is even possible under well chosen conditions \cite{at_b}. In this article, we give a detailed explanation of this result. We present a general, non-perturbative study of the interaction between a cold atomic cloud and quasi-resonant light in the presence of a magnetic field.

The paper is organized as follows. In section \ref{sec_scat}, we derive the scattering operator of a single atom exposed to monochromatic quasi-resonant light in the presence of a magnetic field $\bB$ and we analyze the differential and total cross-section. In section \ref{sec_prop}, we address the coherent propagation of light and derive the refraction index tensor of the effective medium. In particular, birefringence and magneto-optical effects are studied in \ref{susec_bif} and \ref{susec_mo}. The CBS interference effect under a magnetic field is discussed in section \ref{sec_cbsb} and the restoration of the CBS contrast is studied in section \ref{sec_rest}. 

\section{Scattering of light by a single atom in the presence of a uniform magnetic field}\label{sec_scat}

\subsection{Physical setting and basic assumptions}
We consider the situation where a gas of identical atoms, with spatial density $n$, is both exposed to a uniform external magnetic field $\bB$ and a monochromatic light plane wave with wavelength $\lambda$, wavevector $\bk$ ($k=2\pi/\lambda$), angular frequency $\omega=k$ (we use units such that $\hbar=c=1$) and transverse polarization $\bs\epsilon$.

We first assume that the average distance between atoms is much larger than the wavelength $\lambda$ of the light wave, $n\lambda^3 \ll 1$, meaning that multiple scattering  processes take place in the dilute regime. In this case a semi-classical description is appropriate and propagation of light in the atomic medium is well described by partial waves propagating along classical multiple scattering paths. Each path consists in a succession of independent scattering events by a {\it single} atom, separated by propagation in an effective medium with refractive index $N_r$.

To further properly describe scattering of light by one single atom, we assume that $\omega$ is very close to the transition frequency $\omega_0$ between the atom internal groundstate (with total angular momentum $F$) and an internal excited state (with angular momentum $F_e$), hereby considering the case of {\it quasi-resonant} scattering. Introducing the light detuning $\delta = \omega-\omega_0$,  this condition reads $|\delta|\ll \omega_0$. We further assume that this transition is {\it closed} (the excited state with life time $1/\Gamma$ can only decay by spontaneous emission into the groundstate) and well {\it isolated} from any other allowed transition connecting the groundstate to any other hyperfine levels. This is experimentally achieved for instance in the case of the $F=3\rightarrow F_e=4$ transition of the D2 line of $^{85}$Rb \cite{StructInt}.

The total atom-light Hamiltonian $H=H_0+U$ is the sum of the free atom-light Hamiltonian $H_0$ and of the interaction term $U$ describing the coupling of the light modes to the atomic degrees of freedom. When no magnetic field is applied, the interaction between light and one single atom is well documented \cite{CohenGris,CohenRouge} and we thus simply need to incorporate the effect of the magnetic field. We will assume here that weak fields are applied to the gas so that the linear Zeeman effect is the relevant physical description. As the magnetic field only affects the energy levels of the atom, the free atom-light Hamiltonian $H_0$ just reads:
\begin{equation}\label{eq_H0}
H_0=\omega_0\hat{P}_e+\mu g\bB\cdot \bs F+\mu g_e \bB\cdot\bs F_e + \sum_{\bk,\bs\epsilon}\omega_\bk a^{\dagger}_{\bk\bs\epsilon}a_{\bk\bs\epsilon}.
\end{equation}
Here  $\mu \approx 1.4$ MHz/Gauss is the Bohr magneton and $g$ and $g_e$ are the Land\'e factors of the ground and excited states respectively. In the following we will choose $\bB$ to be along axis $0z$ which we choose as the quantification axis for the internal Zeeman states of the atom. Then $\hat{P}_e=\sum_{m_e=-F_e}^{F_e}\ket{F_em_e}\bra{F_em_e}$ is the projector onto the excited state where $\ket{F_em_e}$ denotes an atomic state of angular momentum $F_e$ and magnetic quantum number $m_e$ referred to axis $0z$. 

To be consistent with our approximations, the Zeeman shifts (of order $\mu B$) must be much smaller than the energy difference between any other atomic hyperfine levels. Still these shifts can be sufficiently large to fully split the Zeeman structure and strongly modify the light scattering properties of the atom. Indeed, as exemplified by the case of Rubidium atoms, Zeeman shifts comparable to the excited-state frequency width ($\Gamma/2\pi = 5.9$ MHz), are achieved with moderate field strengths of $5$ Gauss. 

As usual $a_{\bk\bs\epsilon}$ and $a^{\dagger}_{\bk\bs\epsilon}$ in (\ref{eq_H0}) are the annihilation and creation operators of the electric field mode with wavevector $\bk$, polarization $\bs\epsilon\perp\bk$ and frequency $\omega_\bk$. We further assume that the light wavelength is much larger than the size of the relevant atom electronic wavefunctions. The interaction between light and an atom located at position $\br$ is then accounted for in the dipolar approximation and reads $U(\br)=-\bs d_e\cdot \bE_\perp(\br)$. The atomic electric dipole operator writes $\bs d_e = \bs d + \bs d^\dag $ where
\begin{equation}
\bs d^\dag = \hat{P}_e \bs d_e \hat{P}_g
\end{equation}
is the operator describing transitions from the ground state to the excited state. Here $\hat{P}_g=\sum_{m=-F}^{F}\ket{Fm}\bra{Fm}$ is the projector onto the groundstate sector with $\ket{Fm}$ an atomic state of angular momentum $F$ and magnetic quantum number $m$ referred to axis $0z$. The transverse radiation field operator reads $\bE_\perp(\br) =\boldsymbol{D}(\br) + \boldsymbol{D}^\dag(\br)$  where:
\begin{equation}
%\bE_\perp(\br)=i\left (\sum_{\bk,\bs\epsilon}\mathcal{E}_{\omega_\bk}\bs\epsilon_\bk a_{\bk\bs\epsilon}e^{i\bk\cdot \br}+\,\textrm{h.c.}\right ).
\boldsymbol{D}(\br)=i \sum_{\bk,\bs\epsilon}\mathcal{E}_{\omega_\bk}\bs\epsilon_\bk a_{\bk\bs\epsilon}e^{i\bk\cdot \br}
\end{equation}
describes photon annihilation in all possible field modes. The field strength is $\mathcal{E}_{\omega_\bk}=\sqrt{\omega_\bk/2\varepsilon_0\mathcal{V}}$, where $\mathcal{V}$ is the quantization box volume (it disappears at the end of all physical calculations).

The dipolar interaction operator $U$ describes the possibility for the atom to absorb or emit a photon changing at the same time its internal state $\ket{Fm}\rightarrow \ket{F_em_e}$ or $\ket{F_em_e}\rightarrow\ket{Fm}$. %Transitions between the sub-levels of the %ground sate, or between the sub-levels of the excited state are negligeable %because their frequency of order $\mu B$ is much smaller than $\omega$. This %means that 
%The terms linear in $B$ in $H_0$  are not coupled to $U$.
As the incident light is nearly resonant with the atomic transition, we only need to consider resonant contributions where the atom in its ground state can only absorb a photon, and the atom in its excited state can only emit a photon:
\begin{equation}\label{eq_U}
U(\br) \approx \boldsymbol{D}(\br) \cdot \bs d^\dag +\boldsymbol{D}^\dag(\br) \cdot \bs d.
\end{equation}
This is known as the rotating wave approximation.

It has to be noticed that the photons considered in this article are associated to the \emph{transverse} electric field, i.e. to the electric displacement vector $\bs D$ and not to the electric field $\bs E$ (hence our notation) \cite{CohenGris}. This point is important for two reasons: first, due to the magnetic field, light propagates in an anisotropic medium where $\bs D$ and $\bs E$ are no longer colinear and, second, former theoretical studies treated the propagation of $\bs E$ \cite{CBSBthBart}.

\subsection{Scattering amplitude}\label{susec_scat}

Starting from the incident state $\ket{i} = \ket{\bk\omega\bs\epsilon;Fm}$, describing a photon $\ket{\bk\omega\bs\epsilon}$ impinging an atom located at $\br$ with initial ground state $\ket{Fm}$, we consider the scattering process leading to the final state $\ket{f} = \ket{\bk'\omega'\bs\epsilon';Fm'}$, describing a scattered photon $\ket{\bk'\omega'\bs\epsilon'}$ leaving the atom in the final ground state $\ket{Fm'}$. The probability amplitude for such a transition is the matrix element $\bra{f}S\ket{i}$ of the scattering operator $S=\mathbbm{1}-2i\pi T$ acting on the atom-photon Hilbert space $\mathcal{H}=\mathcal{H}_{at}\otimes\mathcal{H}_L$. The transition operator for an atomic point-dipole scatterer writes
\begin{equation}
T = \frac{1}{2\mathcal{V}} \ \ket{\br} \! \bra{\br} \otimes \mathcal{T}
\end{equation}
where $\mathcal{T}$ couples the photon polarization to the atomic internal degrees of freedom. The (on-shell) matrix elements of $T$ are:
%Starting from an incident photon $\ket{\bk\omega\bs\epsilon}$ impinging an atom with initial ground state $\ket{Fm_i}$, we consider the scattering process  leading to a scattered photon $\ket{\bk'\omega'\bs\epsilon'}$ and the atom left in the final ground state $\ket{Fm_f}$. The probability amplitude for such a transition between the initial total state $\ket{\psi_i} = \ket{\bk\omega\bs\epsilon;Fm_i}$ and the total final state $\ket{\psi_f} = \ket{\bk'\omega'\bs\epsilon';Fm_f}$ is the matrix element $\bra{\psi_f}S\ket{\psi_i}$ of the scattering operator $S=\mathbbm{1}-2i\pi T$ acting on the atom-photon Hilbert space $\mathcal{H}=\mathcal{H}_{at}\otimes\mathcal{H}_L$. $T$ defines the transition operator and its matrix elements are
\begin{equation}
%\bra{f}T\ket{i}=\delta(\omega-\omega'+\mu gB(m_i-m_f)) \  \mathcal{T}_{fi}
\bra{f}T\ket{i}=\frac{1}{2\mathcal{V}} \ e^{i(\bk-\bk')\cdot\br} \ \delta(E'-E) \  \mathcal{T}_{fi}(E)
\end{equation}
where the delta distribution ensures energy conservation. For the initial and final states under consideration, $E' = \omega'+\mu gBm'$ and $E = \omega+\mu gBm$. It is important to note at this point that, contrary to the case where there is no magnetic field, scattering has now become {\it inelastic}, a feature that will complicate greatly the analysis of the multiple scattering situation. Indeed as soon as $m' \neq m$, the atom experiences a net Zeeman energy change. In turn, the angular frequency of the scattered photon is accordingly modified to $\omega'=\omega+\mu gB(m-m')$. It should be noted that we simplified here the problem further by neglecting momentum transfer to the atom during scattering. This approximation, supposedly valid when the recoil energy is negligible compared to the energy width of the excited state, becomes nevertheless questionable when considering the multiple scattering regime as studied (without magnetic field) in \cite{DynBD}.

The matrix element $\mathcal{T}_{fi}(E)$ is calculated from $\mathcal{T}_{fi}(E)=\bra{f} UG_e(E)U\ket{i}$ where $G_e(E)=\hat{P}_e(E-H)^{-1}\hat{P}_e$ is the Green's function of the system when the atom is in its excited state. This operator can be computed by using resolvent techniques \cite{CohenRouge} and we find:
\begin{equation}
G_e(E) = \sum_{m_e=-F_e}^{F_e} \frac{\ket{F_em_e}\bra{F_em_e}}{E-g_em_e\mu B - \omega_0 +i \Gamma/2},
\end{equation}
where $\Gamma$ is the angular width of the atomic excited state due to coupling to vacuum fluctuations. Technically speaking, there is also a modification of the atomic frequency (Lamb-shift) that we get rid off by a proper re-definition of $\omega_0$. It is noteworthy that neither the Lamb-shift nor the linewidth depend on $\bB$. This is so because the Zeeman operators do not couple to the dipole interaction $U$.

Introducing the reduced atomic dipole operator $\hat{{\bf d}}_e = {\bf d}_e/d_e$ with $d_e^2\omega_0^3 = 3\pi\epsilon_0 c^3\hbar\Gamma$ \cite{CohenRouge}, the internal transition matrix element then reads:
%\begin{widetext}
\begin{equation}\label{eq_T}
%\bra{\bk'\omega'\bs\epsilon',Fm_f}T%(\omega+\mu gBm_i+i0^+)
%\ket{\bk\omega\bs\epsilon,Fm_i}\!\!
%\mathcal{T}_{fi}(E)=\frac{6\pi}{k^2} \ \frac{\Gamma/2}{\delta+i\Gamma/2} \ t_{m'm}(\bs\epsilon',\bs\epsilon).
\mathcal{T}_{fi}(E)=\frac{6\pi}{k^2} \ \frac{\Gamma/2}{\delta+i\Gamma/2} \ \overline{\bs\epsilon'}\cdot t_{m'm} \cdot\bs\epsilon,
\end{equation}
%\end{widetext}
where the dyadic transition operator in polarization space $t_{m'm}$ is:
%through $t_{m'm}(\bs\epsilon',\bs\epsilon)=\overline{\bs\epsilon'}\cdot t_{m'm} \cdot\bs\epsilon$, we have:
\begin{equation}\label{eq_t}
%t_{m'm}(\bs\epsilon',\bs\epsilon)=\sum_{m_e=-F_e}^{F_e}\frac{\bra{Fm'}(\hat{{\bf d}}_e \cdot \overline{\bs\epsilon'})\ket{F_em_e}\bra{F_em_e}(\bs \hat{{\bf d}}_e\cdot \bs\epsilon)\ket{Fm}}{1-i\phi(gm-g_em_e)}
t_{m'm} = \sum_{m_e=-F_e}^{F_e}\frac{\bra{Fm'}\hat{{\bf d}}_e\ket{F_em_e}\bra{F_em_e}\bs \hat{{\bf d}}_e\ket{Fm}}{1-i\phi(gm-g_em_e)}.
\end{equation}
The dimensionless parameter
\begin{equation}\label{eq_phi}
\phi=\frac{\phi_B}{1-2i\delta/\Gamma},
%\hspace{1cm}  \phi_B = \frac{2\mu B}{\Gamma}.
\end{equation}
where
\begin{equation}\label{eq_phiB}
\phi_B = \frac{2\mu B}{\Gamma},
\end{equation}
quantifies the impact of $\bB$ on the atomic scattering properties and is generally complex-valued except at resonance ($\delta=0$). One can note in passing that, when $B=0$, $t_{m'm}=\bra{Fm'}\bs d \bs d^\dag\ket{Fm}$ which describes the usual absorption and emission cycle from the degenerate groundstate \cite{StructInt2}.

\subsection{Scattering cross section}\label{susec_scatcross}
Probability conservation assures that, under the scattering process, any incident photon is either transmitted in the same mode or scattered in another mode. This is the essence of the optical theorem, which relates the total scattering cross section of the atom to the forward scattering amplitude:
\begin{equation}\label{eq_optheo}
\sigma = -2\mathcal{V} \, \textrm{Im} \bra{\bk\omega\bs\epsilon}\mv{T}_{int}\ket{\bk\omega\bs\epsilon}
\end{equation}
where $\mv{\bullet}_{int}=\textrm{Tr}(\bullet \ \rho_{at})$ indicates an average over the initial atomic internal degrees of freedom. As such $2\mathcal{V}\mv{T}_{int} = \ket{\br} \! \bra{\br} \otimes \mv{\mathcal{T}}_{int}$ only acts on the photon degrees on freedom. At this point, we make the important simplifying assumption that the initial atomic density operator $\rho_{at}$ describes a complete incoherent mixture of the groundstate Zeeman sub-states. This is generally the situation for cold atoms prepared in an optically thick magneto-optical trap (MOT) where all Zeeman states are uniformly populated and no Zeeman coherence is achieved. The initial atomic density operator then reads $\rho_{at}=\hat{P}_g/(2F+1)$.
%some straightforward algebra leads to 
%\begin{equation}\label{eq_tint}
%\mv{T}_{int} = \frac{1}{2\mathcal{V}} \ \ket{\br} \!\! \bra{\br} \otimes \mv{t}
%\mv{T}_{int}=M_{FF_e}\frac{3\pi e^{i(\bk-\bk')\cdot\br}}{\mathcal{V}\omega^2}\frac{\Gamma/2}{\delta+i\Gamma/2}\Delta_{\bk'}\calTn{FF_e}\Delta_\bk 
%\end{equation}
Using the fact that $\hat{{\bf d}}_e$, $\bs\epsilon$ and $\overline{\bs\epsilon'}$ are irreducible tensors of rank 1, one can express $\overline{\bs\epsilon'}\cdot\mv{t}_{int}\cdot \bs\epsilon$ in terms of the irreducible components of the tensor $\overline{\bs\epsilon'}_i\bs\epsilon_j$ in the cartesian basis set $(\hat{\bx}, \hat{\by}, \hat{\bz})$. Introducing the dyadic projector $\Delta_{\bs a} = \mathbbm{1} - \bs a \bs a /a^2$ onto the plane perpendicular to $\bs a$, we find:
\begin{eqnarray}\label{eq_tint}
\mv{\mathcal{T}}_{int} &=& M_{FF_e}\frac{6\pi}{k^2}\frac{\Gamma/2}{\delta+i\Gamma/2}\Delta_{\bk'}\calTn{FF_e}\Delta_\bk \\
\calTn{FF_e}&=& \frac{3}{2F_e+1}\sum_{m=-F}^{F} t_{mm},
\end{eqnarray}
where $M_{FF_e}=(2F_e+1)/(3(2F+1))$ is a factor taking care of the degeneracies of the ground and excited states. Both projectors $\Delta_\bk$ and $\Delta_{\bk'}$ ensure that polarization vectors always remain transverse to the direction of propagation.

Using the totally antisymmetric tensor of rank 3 $\varepsilon_{ijk}$ ($\varepsilon_{xyz}=1$), the dyadic tensor $\calTn{FF_e}$ reads:
\begin{equation}\label{dyadicT}
\calTn{FF_e}_{ij}=\zeta\delta_{ij}+\eta\varepsilon_{ijk}\hat{\bB}_k+\xi \hat{\bB}_i\hat{\bB}_j.
\end{equation}
With the z-axis chosen along $\hat{\bB}$, the matrix representing $\calTn{FF_e}$ is:
\begin{equation}\label{eq_TF}
\calTn{FF_e}= \left (
\begin{array}{ccc}
\zeta & \eta & 0\\
-\eta&\zeta&0\\
0&0&\zeta+\xi
\end{array} \right ).
\end{equation}
The $\zeta$, $\eta$ and $\xi$ coefficients depend on $\phi$ and on the Clebsch-Gordan coefficients of the atomic transition \cite{Edmonds,Messiah}:
\begin{eqnarray}\label{eq_coefs}
\zeta&=&\frac{1}{2}\bigg (\sum_{m=-F}^{F}\frac{\langle
  FF_e-mm+1|11\rangle^{2}}{1-i\phi((g-g_e)m-g_e)}\nonumber\\
&&\qquad\qquad+\frac{\langle
  FF_e-mm-1|1-1\rangle^{2}}{1-i\phi((g-g_e)m+g_e)}\bigg )\nonumber\\
\eta&=&-\frac{i}{2}\bigg (\sum_{m=-F}^{F}\frac{\langle
  FF_e-mm+1|11\rangle^{2}}{1-i\phi((g-g_e)m-g_e)}\nonumber\\
&&\qquad\qquad-\frac{\langle
  FF_e-mm-1|1-1\rangle^{2}}{1-i\phi((g-g_e)m+g_e)}\bigg )\nonumber\\
\xi&=&-\zeta+\sum_{m=-F}^{F}\frac{\langle
FF_e-mm|10\rangle^{2}}{1-i\phi(g-g_e)m}
\end{eqnarray}
As is expected from Onsager's reciprocity relations \cite{Onsager}, one can check that $\zeta$ and $\xi$ are even function of $\phi$, while $\eta$ is an odd function of $\phi$. It can be checked that all these three coefficients are real at resonance ($\delta=0$) since $\phi=\phi_B$ is then real. In the case of a $F=0\rightarrow F_e=1$ transition ($g=1$), these coefficients read
\begin{eqnarray}\label{CoeF=0}
\zeta&=&\frac{1}{1+(g_e\phi)^2}\nonumber\\
\eta&=&-\frac{g_e\phi}{1+(g_e\phi)^2}\\
\xi&=&\frac{(g_e\phi)^2}{1+(g_e\phi)^2}\nonumber
\end{eqnarray}

The dyadic tensor $\calTn{FF_e}$ embodies the effect of the magnetic field on the photon polarization degrees of freedom and gives rise to the usual magneto-optical effects. The $\zeta$ term is responsible for normal extinction (Lambert-Beer law). The $\eta$ term describes the magnetically-induced rotation
of the atomic dipole moment (Hanle effect) \cite{Hanle,PHE} and induces Faraday rotation and dichroism effects observed when $\bm{k} \parallel \bB$ \cite{Landau}. The $\xi$ term is responsible for the Cotton-Mouton effect observed when $\bm{k} \perp \bB$ \cite{Landau}.\\

From expression (\ref{eq_tint}) and the optical theorem (\ref{eq_optheo}) we deduce the total scattering cross section of a photon by an atom initially prepared in an incoherent mixture of Zeeman internal ground states:
\begin{eqnarray}\label{eq_sigma}
\sigma(\phi)&=&\sigma \ \textrm{Re}\left( (1+2i\frac{\delta}{\Gamma})\ (\overline{\bs\epsilon}\cdot\calTn{FF_e}\cdot\bs\epsilon) \right)\\
&=&\sigma \textrm{ Re }\left ((1+2i\frac{\delta}{\Gamma})(\zeta+\eta(\overline{\bs\epsilon}\times\bs\epsilon)\cdot\hat{\bs B}+\xi|\bs\epsilon\cdot\hat{\bs B}|^2\right )\nonumber
\end{eqnarray} 
where 
\begin{equation}
\sigma = \frac{\sigma_0}{1+(2\delta/\Gamma)^2} \hspace{1cm} \sigma_0 = M_{FF_e} \ \frac{6\pi}{k^2}.
\end{equation}
In the case of a $F=0\rightarrow F_e=1$ transition ($g=1$), and for resonant light ($\delta=0$), the scattering cross section boils down to the simpler form:
\begin{equation}
\sigma(\phi)=\sigma_0\frac{1+(g_e\phi_B)^2|\bs\epsilon\cdot\hat\bB|^2}{1+(g_e\phi_B)^2}.
\end{equation}
As one can see, in the presence of a magnetic field, the total scattering cross section depends explicitly on the incident polarization $\bs\epsilon$. More precisely, it depends on the relative direction of $\bs\epsilon$ with respect to $\bB$. We recover here the well-known fact that an external magnetic field induces optical anisotropy in otherwise isotropic media. In the absence of a magnetic field, $\calTn{FF_e}$ reduces to the identity matrix $\mathbbm{1}$ and we get $\sigma(\phi =0)=\sigma$, giving back the result found in \cite{StructInt2}.

\subsection{Impact of optical pumping}\label{susec_po}
Our previous calculation in fact just considered the scattering of {\it one} quasi-resonant photon by one atom assumed to be initially prepared in an incoherent mixture of Zeeman internal ground states. In a real experiment however, many quasi-resonant photons are shone. For a given incident polarization, the repeated scattering of photons by the atom induces changes in the populations of the groundstate Zeeman sublevels and creates coherences between them. This effect is known as optical pumping and is enhanced in the presence of a magnetic field. Our previous calculations thus applies to the case where optical pumping can be neglected. This is the situation considered in section \ref{sec_rest}. One can note however that our results can be easily extended to include optical pumping. Indeed, the most general tensor of rank 2 that can be written with the components of $\bB$ is still given by expression (\ref{dyadicT}) but with arbitrary coefficients now. Thus equations (\ref{eq_tint}), (\ref{eq_TF}) and (\ref{eq_sigma}) remain valid. Only the detailed expressions (\ref{eq_coefs}) of $\zeta$, $\eta$ and $\xi$ will be modified. In the following, all our analytical results are expressed in terms of these three coefficients. They thus remain valid in the presence of optical pumping if the coefficients are given their appropriate expressions and values.

\subsection{Differential scattering cross section}\label{susec_dscat}
We now turn to the impact of $\bB$ on the radiation pattern of the atom. The intensity of light scattered in the direction $\bk'$ with polarization $\bs\epsilon'$ while the atom changes its internal state from $\ket{Fm}$ to $\ket{Fm'}$ is proportional to the differential scattering cross section
\begin{eqnarray}
\frac{d\sigma_{m'm}}{d\Omega}(\bk\bs\epsilon\rightarrow\bk'\bs\epsilon')&=&\frac{\mathcal{V}^2\omega^2}{(2\pi)^2}|\bra{\bk'\bs\epsilon',Fm'}T\ket{\bk\bs\epsilon, Fm}|^2 \nonumber \\
&=& \frac{9\sigma}{8\pi}\ \frac{2F+1}{2F_e+1} \ |\overline{\bs\epsilon'}\cdot t_{m'm} \cdot\bs\epsilon|^2
\end{eqnarray}
Starting with an atom prepared in state $\ket{Fm}$ the differential cross-section for a photon to be scattered in mode $\ket{\bk'\bs\epsilon'}$ is thus:
\begin{equation}
\frac{d\sigma_{m}}{d\Omega}(\bk\bs\epsilon\rightarrow\bk'\bs\epsilon')=\sum_{m'=-F}^{F} \frac{d\sigma_{m'm}}{d\Omega}(\bk\bs\epsilon\rightarrow\bk'\bs\epsilon')
\end{equation}
The total photon scattering cross-section is then obtained by averaging over the initial internal atomic density matrix, which we assumed to describe a fully incoherent mixture of groundstate Zeeman sublevels. We thus arrive at:
\begin{eqnarray}\label{eq_avdsec}
\frac{d\sigma}{d\Omega}(\bk\bs\epsilon\rightarrow\bk'\bs\epsilon')&=&\frac{1}{2F+1}\sum_{m,m'=-F}^{F}\frac{d\sigma_{m'm}}{d\Omega}(\bk\bs\epsilon\rightarrow\bk'\bs\epsilon') \nonumber \\
&=& \frac{9\sigma}{8\pi} \ \sum_{m,m'=-F}^{F} \frac{|\overline{\bs\epsilon'}\cdot t_{m'm} \cdot\bs\epsilon|^2}{2F_e+1}.
%\frac{|t_{m'm}(\bs\epsilon',\bs\epsilon)|^2}{2F_e+1}.
\end{eqnarray} 
When the ground state is non-degenerate ($F=0$, $F_e=1$, $g=1$), the tensors $t_{00}$ and $\calTn{01}$ coincide and the differential cross section takes the simple form: 
\begin{equation}
\frac{d\sigma}{d\Omega}(\bk\bs\epsilon\rightarrow\bk'\bs\epsilon')= \frac{3\sigma}{8\pi}|\overline{\bs\epsilon'}\cdot \calTn{01}\cdot\bs\epsilon|^2,
\end{equation}
with
\begin{equation}
\overline{\bs\epsilon'}\cdot \calTn{01}\cdot\bs\epsilon = \zeta \ \overline{\bs\epsilon'}\cdot\bs\epsilon+\eta \ (\overline{\bs\epsilon'}\times\bs\epsilon)\cdot\hat{\bs B}+ \xi \ (\overline{\bs\epsilon'}\cdot\hat{\bs B})(\overline{\bs\epsilon}\cdot\hat{\bs B}),
\end{equation}
the coefficients $\zeta$, $\eta$ and $\xi$ being given by (\ref{CoeF=0}).

Since the resonant denominator in (\ref{eq_t}) explicitly depends on the magnetic quantum numbers $m$ and $m_e$, standard irreducible tensor methods \cite{StructInt2} are of little practical use to boil down the total differential cross-section (\ref{eq_avdsec}) to a much simpler form. However it can be easily and efficiently computed for any transition line by using the symbolic calculus software \textit{Maple}\texttrademark. As a humorous note, the reader is invited to appreciate the power of the optical theorem by deriving the total scattering cross section (\ref{eq_sigma}) by direct computation from (\ref{eq_avdsec}). \\

\section{Coherent propagation of light}\label{sec_prop}
The propagation of an incident light field mode $\ket{\bk\bs\epsilon}$ in a scattering medium, also known as the coherent propagation, is efficiently described by replacing the scattering medium by a homogeneous effective medium having a complex refractive index tensor $N_r$. When $B=0$, $N_r$ is a scalar and bears no action on the incident polarization. Its imaginary part gives rise to an exponential attenuation (Lambert-Beer law), its characteristic length scale being known as the extinction length. Since true absorption (i. e. conversion of electromagnetic energy into another form of energy) is absent in our case, depletion of the incident mode can only occur through scattering. The extinction length thus identifies with the scattering mean free path $\ell$ and one has $\textrm{Im}(N_r) = 1/(2k\ell)$.

The presence of a magnetic field $\bB$ modifies the interaction between light and atoms and the coherent propagation of light in the atomic cloud will be accordingly altered. As $\bB$ induces a preferential orientation of space, it will set an optical anisotropy in the atomic gas. In this case, the refractive index tensor is no longer a scalar and will act on the polarization space: the coherent propagation of light will exhibit birefringence and magneto-optical effects. These magneto-optical effects are well documented in the literature \cite{CBSBthMartinez,CBSBthBart,VoigtCold}: for a given wavevector $\bk$, there are two eigen-polarization modes propagating in (possibly) different directions, with different velocities and attenuations. 

In this paragraph, we extend the techniques developed in \cite{CBSBthBart} and \cite{AdBart} to the case of atoms with a degenerate ground state. We will always assume that the atomic gas is dilute. In this case, $n\lambda^3\ll 1$ ($n$ being the number density), and $\ell = 1/(n\sigma)$.  Furthermore, since for resonant scatterers $\sigma$ is at most of the order of $\lambda^2$, $\ell$ will be always much larger than the average interatomic distance $n^{-1/3}$. As a whole, for point-dipole resonant scatterers, the dilute medium condition $n\lambda^3\ll 1$ implies the condition $k\ell \gg 1$. The properties of the effective medium will then be directly related to the individual scattering properties of the atoms under the magnetic field.

\subsection{Average Green's function for light propagation}
The first step to find the refractive index tensor $N_r$ and describe the coherent propagation of light is to determine the Green's function for light once all atomic degrees of freedom have been averaged out.

Let $N$ be the total number of atoms in the gas and their respective positions be labeled by $\bs r_i$ ($i=1, \cdot\cdot\cdot N$). The Hamiltonian of the system $\{$atoms + light$\}$ is
\begin{eqnarray}\label{eq_Htot}
\mathcal{H}&&=\sum_{i=1}^{N}H_0(\br_i)+\sum_{i=1}^{N}U(\br_i)\nonumber\\
&&=\mathcal{H}_0+\mathcal{U},
\end{eqnarray}
where $H_0$ and $U$ are given by expressions (\ref{eq_H0}) and (\ref{eq_U}). The Green's functions of the whole system $G(z)=(z-\mathcal{H})^{-1}$ and of the uncoupled system $G_0(z)=(z-\mathcal{H}_0)^{-1}$ satisfy the recursive equation 
\begin{eqnarray}
G(z)&=&G_0(z)+ G_0(z) \mathcal{U}G(z) \nonumber \\
&=& G_0(z)+G_0(z)\mathcal{U}G_0(z)+\ldots
\end{eqnarray}
Since we are interested in the situation where all atoms start and end in their groundstate, we merely look for the Green's function projected onto the atomic groundstate manifold $\hat{P}_gG(z)\hat{P}_g$. In this case, only expansion terms containing $\mathcal{U}$ an even number of times can contribute. %Introducing the transition operator $T_i(z)=U(\br_i)+U(\br_i)G_0(z)U(\br_i)+\ldots$ of the $i$-th atom,
The average over the atomic degrees of freedom will generate a Dyson equation for the average Green's function $\mathcal{G}(z)= \langle \hat{P}_gG(z)\hat{P}_g \rangle$. Under the dilute medium assumption, and since all atoms are identical and uniformly distributed in space (at least on the scale of the scattering mean free path), the corresponding self-energy is given by $\Sigma(z) = N \langle T_i(z)\rangle_{int}$ where $T_i(z)=U(\br_i)+U(\br_i)G_0(z)U(\br_i)+\ldots$ is the transition operator of the $i$-th atom. The average over the internal degrees of freedom is given by (\ref{eq_tint}).

The matrix elements of the average photon Green's function are then given by:
\begin{eqnarray}
\langle \bk' \bs\epsilon' | \mathcal{G}(\omega) | \bk\bs\epsilon\rangle &=& \delta_{\bk\bk'} \ \overline{\bs\epsilon'}\cdot \mathcal{G}(\bk,\omega)\cdot\bs\epsilon \label{eq_Gk1}\\
\mathcal{G}(\bk,\omega)&=&\Delta_\bk \frac{1}{\omega-k-\Sigma(k, \hat{\bk})} \Delta_\bk \label{eq_Gk} \\
\Sigma(k,\hat{\bk})&=&\frac{1}{2\ell_0}\frac{\Gamma/2}{\delta+i\Gamma/2}\Delta_\bk\calTn{FF_e}\Delta_\bk
%\mathcal{G}(\bk,\omega)=\frac{1}{\mv{G_{0}}_{int}^{-1}-\Sigma(k,\hat{\bk})}=\frac{1}{\omega-k-\Sigma(k,\hat{\bk})}
\label{eq_S}
\end{eqnarray}
where $\ell_0=1/(n\sigma_0)$. The Kronecker symbol $\delta_{\bk\bk'}$ features the restoration of translation invariance under the spatial average. The self-energy tensor $\Sigma(k,\hat{\bk})$ contains all the information on the effective medium. It has an explicit dependence on the incident direction because, in the presence of the magnetic field, the scattering medium develops an optical anisotropy (see section \ref{susec_scatcross}). 

\subsection{Optical anisotropy}\label{susec_anisop}
The equation (\ref{eq_Gk}) receives a simple interpretation in a basis where $\Sigma(k,\hat{\bk})$ is diagonal. It corresponds to polarization modes which propagate in direction $\hat{\bk}$ without deformation. The poles of $\mathcal{G}(\bk,\omega)$ then give the corresponding dispersion relation for the eigenmode.

Trivially, $\bk$ is an eigenvector of $\Sigma(k, \hat{\bk})$ with eigenvalue 0. This is a consequence of the transversality of light, and this eigenvector is not physically relevant. To find the other complex eigenmodes and complex eigenvalues,  $\Sigma(k, \hat{\bk})\hat{\bs V}_{\pm}=\Lambda_{\pm} \hat{\bs V}_{\pm}$, we parametrize $\hat{\bk}$ by its spherical angles $\theta$ and $\varphi$ in a coordinate frame with z-axis parallel to $\hat{\bB}$. We find:
 \begin{equation}\label{eq_valpm}
\Lambda_{\pm}(\hat{\bk})=\zeta+\xi\frac{\sin^{2}\theta}{2}\pm\sqrt{-\eta^{2}\cos^{2}
\theta+\xi^{2}\frac{\sin^{4}\theta}{4}}
\end{equation}
Their explicit dependance on the angle $\theta$ between $\hat{\bB}$ and $\hat{\bk}$ expresses the optical anisotropy of the atomic cloud induced by the magnetic field. The corresponding eigenmodes are:
%\begin{widetext}
\begin{eqnarray}\label{eq_vecpm}
\hat{\bs V_{\pm}}(\hat{\bk})&\propto&\Big(\eta\cos^{2}\theta\cos\varphi+\xi\frac{\sin^{2}\theta}{2}\sin\varphi\nonumber\\
&&\qquad\qquad\mp\sin\varphi\sqrt{-\eta^{2}\cos^{2}\theta+\xi^{2}\frac{\sin^{4}\theta}{4}} \ \Big) \hat{\bs x}\nonumber\\
&&+\Big(\eta\cos^{2}\theta\sin\varphi-\xi\frac{\sin^{2}\theta}{2}\cos\varphi\nonumber\\
&&\qquad\qquad\pm\cos\varphi\sqrt{-\eta^{2}\cos^{2}
\theta+\xi^{2}\frac{\sin^{4}\theta}{4}} \ \Big) \hat{\bs y}\nonumber\\
&&-\eta\cos\theta\sin\theta \hat{\bs z}
\end{eqnarray}
%\end{widetext}
To allow a relative ease of reading, these vectors have not been normalized.

One difficulty arises when $(-\eta^{2}\cos^{2}\theta+\xi^{2}(\sin^{4}\theta)/4)=0$ for non-vanishing $\eta$ and $\xi$. This situation can only happen if $\eta$ and $\xi$ are real, thus for $\phi$ real, i. e. at resonance ($\delta=0$). The solutions are of the form ($\pm\theta_0, \pi\pm\theta_0$), which means, because of the invariance around the z-axis, that the photon has to propagate along a cone of apex $\theta_0$. When this is the case, $\Lambda_+=\Lambda_-$ but $\hat{\bs V}_+=\hat{\bs V}_-$, and the self-energy is non-diagonalizable. However this situation is largely unphysical in the sense that it arises from the first-order approximation in the atomic density used to compute $\Sigma(k,\hat{\bk})$. At next order in density, this difficulty disappears. However, this could lead to intersting effects for the propagation of light near the apex angle. We chose to neglect such effects in the following.

Noticeably, $\hat{\bs V}_+$ and $\hat{\bs V}_-$ are not orthogonal vectors in general. Indeed, scattering depletes the coherent mode and its energy decays. This is reflected by the fact that the self-energy is not a hermitian operator. Thus its eigenvectors have no reason to be orthogonal. When $B=0$, $\calTn{FF_e}$ reduces to the identity and $\Sigma(k,\hat{\bk})$ is then proportional to the projector $\Delta_\bk$. In this case alone, all polarization states orthogonal to $\bk$ are eigenmodes and it is then possible to choose an orthogonal basis. One can however note that $\hat{\bs V}_+$ and $\hat{\bs V}_-$ are nearly orthogonal when $|\eta|\ll |\xi|$ or when $|\eta|\gg |\xi|$. This happens in the limit of small or strong magnetic fields ($\mu B\ll \delta,\Gamma$, $\mu B\gg\delta,\Gamma$), or at a very large detuning ($\delta\gg\mu B,\Gamma$).

 \subsection{Refraction index }\label{susec_index}
 
In the polarization eigenbasis, the poles of the Green's function (\ref{eq_Gk}) give the dispersion relation for $\hat{\bs V}_\pm$. We find:
\begin{equation}\label{eq_disp}
\omega_\pm(\bk)=k+\frac{1}{2\ell_0}\frac{\Gamma/2}{\delta+i\Gamma/2}\Lambda_\pm(\hat{\bk})
\end{equation}
%In the first order in the atomic density, on finds $\omega(\bk)$ by replacing $\omega$ by $k$ in the expressions of $\delta$ and $\ell_0$.
The refraction index tensor $N_r$ is diagonal in the polarization eigenbasis and the polarization vectors $\hat{\bs V}_\pm$ propagate each with different complex refractive indexes:
%$n_r=k/\omega(\bk)$:
\begin{equation}\label{eq_ind}
N_{r}^{\pm}(\bk)=\frac{k}{\omega_\pm(\bk)} \approx 1-\frac{1}{2k\ell_0}\frac{\Gamma/2}{\delta+i\Gamma/2}\Lambda_{\pm}(\hat{\bk})
\end{equation}
since $k\ell_0 \gg 1$. As a consequence, the two eigen-polarizations propagate with different phase velocities and experience different attenuations (dichroism). In turn, the effective medium acts as an absorbing polarization filter for the incoming light.

The index mismatch between the two eigen-polarizations is
\begin{eqnarray}
\Delta N_r = N_{r}^{+}-N_{r}^{-} = \frac{\mathrm{i}}{2k\ell_0} \, \frac{1}{1-2\mathrm{i}\delta/\Gamma} \, (\Lambda_{+}-\Lambda_{-})\\ \label{eq_IndMis}
= \frac{\mathrm{i}}{2k\ell_0} \, \frac{1}{1-2\mathrm{i}\delta/\Gamma} \, \sqrt{-\eta^2\cos^2\theta+\xi^2\frac{\sin^4\theta}{4}}
\end{eqnarray}
and vanishes for $\theta=\theta_0$, i. e. when the two eigen-polarizations collapse onto each other, rendering the refractive index tensor no longer diagonalizable.

\subsection{Group velocity and birefringence}\label{susec_bif}

Let us consider a polarized monochromatic wave packet propagating in the atomic cloud with a central wavevector $\bk$, and let us assume the polarization is one of the vectors $\hat{\bs V}_\pm$. Then the maximum of the wave packet propagates with the group velocity $\bv_g=\textrm{Re}(\bnabla_\bk\omega(\bk))$. As $\omega(\bk)$ depends on the angle $\theta$ between $\bB$ and $\bk$, $\bv_g$ possesses a component orthogonal to $\bk$: in general, the wave packet does not propagate parallel to the wavevector. Since each polarization eigen-mode has its own direction of propagation, a birefringence effect takes place, as is well known in anisotropic media \cite{BW}. This magnetically-induced birefringence has already been observed \cite{Schlesser}, though under conditions differing from the ones described in the present article.

We have checked that, provided $|\delta|$ is not much larger than $\mu B$, the deviation of the wave packet from the direction of $\bk$ is negligible: the walk-off angle remains smaller than $0.05/(k\ell_0)$ \cite{These}. Birefringence effects will thus be neglected in the following. 

\subsection{Propagation in real space}\label{susec_propre}

To study the propagation properties of the coherent mode in the atomic cloud, we need the Green's function of light  in real space, $G_\omega(\br)$. It is the Fourier transform of $\mathcal{G}(\bk,\omega)$ (\ref{eq_Gk}):
\begin{equation}
G_\omega(\br)=\int \frac{d^3\bk}{(2\pi)^3}\, \Delta_\bk \, \frac{e^{i\bk\cdot\br}}{\omega-k-\Sigma(k,\hat{\bk})}\Delta_\bk 
\end{equation} 
Neglecting birefringence effects, 
the angular part of the integral can be calculated with a stationary phase approximation around the direction $\hat{\br}$:
\begin{eqnarray}\label{eq_Grrad}
%G(\br,\omega)\!\!=\!\!\frac{1}{(2\pi)^2 ir}\int_0^\infty\!\! dk\frac{k(e^{ikr}-e^{-ikr})}{\omega-k-\frac{\Gamma/2}{2\ell_0(\delta+i\Gamma/2)}\Delta_\br\calTn{FF_e}\Delta_\br}\Delta_\br 
G_\omega(\br) &=& \frac{1}{2\pi^2r} \, \Delta_\br \, I_\omega(\br) \, \Delta_\br \\
I_\omega(\br) &=& \int_0^\infty  \frac{k\sin(kr) \, dk}{\omega-k-\frac{\Gamma/2}{2\ell_{res}(\delta+i\Gamma/2)}\Delta_\br\calTn{FF_e}\Delta_\br}
\end{eqnarray}
The $I_\omega(\br)$ integral has an ultra-violet divergence ($k\to\infty$) which needs to be regularized. It expresses that the interaction between two atoms separated by less than one optical wavelength cannot be reduced to the exchange of one resonant photon. The divergent part of the integral comes from the contact term of the radiated field \cite{CohenGris}. As the average interatomic distance is much larger than the optical wavelength, we are interested only in the far-field component. The latter can be obtained from the regular part of (\ref{eq_Grrad}) and calculated with the help of the residue theorem. The final result reads:
\begin{equation}\label{eq_Gr}
G_\omega(\br)=-\frac{\omega }{2\pi r}\, \Delta_\br \, e^{ikrN_r}\Delta_{\br}
\end{equation}
featuring the refractive index tensor
\begin{equation}\label{eq_N}
N_r(k,\hat{\br})=\bs 1-\Delta_\br\frac{\Sigma(k,\hat{\br})}{k}\Delta_\br.
\end{equation}
In the presence of a magnetic field, $N_r$ is represented by a matrix with an antisymmetric part proportional to $\eta$, as can be seen from (\ref{eq_S}) and (\ref{eq_TF}). This implies that $G_\omega(\br)$ is not a symmetric operator: the transposition operation is equivalent to flip the sign of $\eta$, or equivalently to flip the sign of $\bB$: 
\begin{equation}
G_\omega(\br, \bB)={}^tG_\omega(\br,-\bB)
\end{equation}
This property is closely related to the reciprocity theorem \cite{RecSaxon}. As a simple illustration, assume a light beam is going successively through a linear polarizer $\bs\epsilon$, the atomic cloud and a linear analyzer $\bs\epsilon'$. The amplitude of the transmitted light is then proportional to $A_{dir}=\bs\epsilon' \cdot G_\omega \cdot \bs\epsilon$. If we now consider the reverse situation where the  light beam is traveling through the system along the opposite direction, its transmitted amplitude will now be $A_{rev}=\bs\epsilon \cdot G_\omega\cdot \bs\epsilon'= \bs\epsilon' \cdot {}^tG_\omega \cdot \bs\epsilon$. We see that $A_{dir}=A_{rev}$ if and only if $G_\omega(\br)$ is symmetric, which is not the case in the presence of the magnetic field. This is in essence the underlying principle behind optical isolators (or optical "diodes"), which are devices realizing $A_{rev}=0$ while $A_{dir} \not= 0$.

\subsection{Magneto-optical effects}\label{susec_mo}

The impact of $\bB$ on the coherent propagation of light is embodied in the refractive index tensor $N_r$ (\ref{eq_N}). As already mentioned in \ref{susec_index}, it concerns a differential dephasing and attenuation of the eigenmodes of propagation. These magneto-optical effects have been extensively studied when light propagates parallel (Faraday effect) or perpendicular to (Cotton-Mouton effect) $\bB$. We present below both effects in terms of our formalism, giving substance to the physical interpretation of the parameters $\zeta$, $\eta$ and $\xi$.

\subsubsection{Faraday effect \label{sususec_farad}}

%\begin{center}
\begin{figure}
\includegraphics[width=7cm]{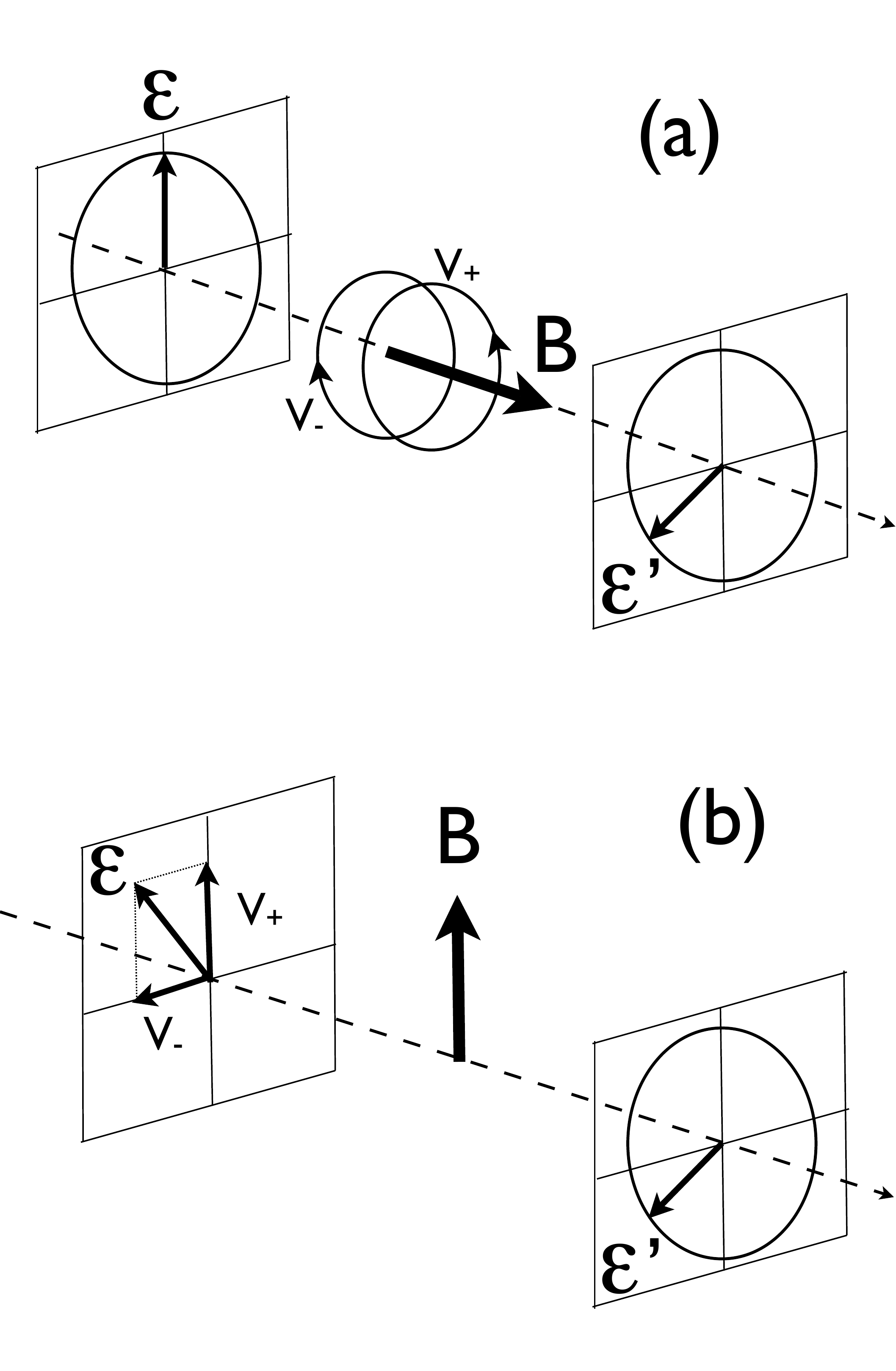}
\caption{\label{fig_farad} (a) Resonant Faraday effect ($\delta=0$). A linearly-polarized light beam propagates parallel to $\bB$. The polarization eigenmodes are the circular polarizations. In the course of propagation the polarization of the light beam keeps linear but rotates around $\bB$ (b) Resonant Cotton-Mouton effect ($\delta=0$). A linearly-polarized light beam propagates perpendicular to $\bB$. The polarization eigenmodes are linear, one being parallel to $\bB$ and the other one being perpendicular to $\bB$. In the course of propagation the polarization of the light beam keeps linear but rotates around the propagation axis.}
\end{figure}
%\end{center}

The Faraday effect occurs when a light beam with a linear polarization propagates along $\bB$ in the atomic cloud. For sake of simplicity, we consider here that the frequency of light is exactly at resonance with the atomic transition ($\delta=0$) so that $\phi=\phi_B$, $\zeta$, $\eta$ and $\xi$ are all real in the following discussion. From (\ref{eq_valpm}) and (\ref{eq_vecpm}), the eigenmodes and their associated eigenvalues are:
\begin{equation}
\hat{\bm V}_{\pm}=(\hat{\bs x}\pm i\frac{|\eta|}{\eta}\hat{\bs y})\qquad
\Lambda_{\pm}=\zeta\pm i|\eta|
\end{equation}

The eigenmodes thus identify with the left and right circular polarization vectors (see Fig.\ref{fig_farad}a). Their index mismatch $\Delta N_r= -|\eta|/(k\ell_0)$ is real, meaning that the two eigenmodes develop a phaseshift in the course of propagation. An elementary calculation shows that the polarization of the traveling beam remains linear but rotates around $\hat{\bB}$ by an angle $\Theta = \eta L/(2\ell_0)$ proportional to the traveled distance $L$.
%$G_\omega(\br)$ acting on a linear polarization will indeed rotate it by the angle $\Theta = \eta r/(2\ell_0)$ around $\hat{\bB}$,
Hence, the parameter $\eta$ describes the Faraday effect. As one can also see, the parameter $\zeta$ plays the same role for the two eigenpolarizations : it acts as an isotropic refractive index.

At low magnetic fields ($\mu B\ll \Gamma$), $\eta$ is proportional to $B$, and so is the rotation angle $\Theta$. The proportionality constant between this angle and the product $BL$ is known as the Verdet constant $V_B$. For the $F=3\rightarrow F_{e}=4$ transition of $^{85}$Rb, one finds 
\begin{equation}
V_B=-\frac{3}{4}\frac{\mu}{\Gamma\ell_0}\sim -8.10^6\,\textrm{rad/(Tm)}
\end{equation}
with $\ell_0\sim 100\mu$m. This value is three orders of magnitude larger than those of classical materials, and comparable to the value measured in \cite{VoigtCold}.

\subsubsection{Cotton-Mouton or Voigt effect}\label{sususec_Voigt}

This effect describes the modification of the polarization of a light beam propagating perpendicularly to $\bB$. As for the Faraday effect, we assume $\delta=0$ to simplify the discussion. The eigenmodes and their eigenvalues now read:
\begin{equation}
\hat{\bs V}_+=\hat{\bs z}\quad \Lambda_+=\zeta+\xi\qquad \hat{\bs V}_-=\hat{\bs y}\quad\Lambda_-=\zeta 
\end{equation}
The eigenmodes are now the linear polarization vectors (see Fig.\ref{fig_farad}b). Their index mismatch $\Delta N_{r} = \mathrm{i} \xi/(2k\ell_0)$ is now purely imaginary, showing that $\hat{\bs V}_+$ gets more attenuated than $\hat{\bs V}_-$ by a factor $e^{-\xi L/(2\ell_0)}$ after traveling the distance $L$. Here again, the parameter $\zeta$ plays the same role for the two propagation eigenmodes and acts as an isotropic refractive index.

If the incident polarization is linear, it can be written as a real linear combination of $\hat{\bs V}_+$ and $\hat{\bs V}_-$. During the propagation, the component along $\hat{\bs V}_+$, i.e. along $\hat{\bB}$, will decrease more than the one along $\hat{\bs V}_-$, i. e. perpendicular to $\hat{\bB}$. As a whole, the polarization of the light beam remains linear, but rotates around $\hat{\bk}$ by an angle which depends on $\xi$, which is thus the parameter describing the Cotton-Mouton effect.

In ref. \cite{BW}, the Cotton-Mouton effect is rather described as the transformation of an incident linear polarization into an elliptical one in the course of propagation, this transformation being a consequence of the accumulated phase shift between the two propagating eigenmodes. This apparent contradiction can be lifted if one realizes that, contrary to our discussion, the Cotton-Mouton effect described in \cite{BW} is in fact the one usually observed for light frequencies which are far-detuned from any resonance frequencies. Work out our theory at a very large detuning $\delta$, we indeed recover the description given in \cite{BW}. 

\subsubsection{General case}

When the direction of propagation $\hat{\bk}$ is neither along nor perpendicular to $\hat{\bB}$, both effects mix. The Faraday effect will dominate when $\hat{\bk}$ is roughly parallel or anti-parallel to $\hat{\bB}$, i. e. essentially when $\hat{\bk}$ is well inside the cone with apex angle $\theta_0$ for which (\ref{eq_IndMis}) vanishes. On the contrary, the Cotton-Mouton effect dominates when $\hat{\bk}$ is essentially well outside this cone, i. e. roughly perpendicular to $\hat{\bB}$. To give orders of magnitude, $\eta$ and $\xi$ are comparable when $\phi\simeq 1$, which corresponds to $B\simeq 2$G for $^{85}$Rb atoms. The apex angle is then $\theta_0 \approx 65^{\circ}$. The Cotton-Mouton effect then dominates for directions of propagation making an angle between 65$^{\circ}$ and 90$^{\circ}$ with $\hat{\bB}$. In classical media, this would happen only in a narrow angular width of order $10^{-4}$rad around the direction orthogonal to $\hat{\bB}$. This big difference in orders of magnitude is due to the strong resonant character of the atoms. All in all, the giant Faraday effect and the large zone of preponderance of the Cotton-Mouton effect make it necessary to take both effects into full account when studying the coherent propagation of light. 

When $\delta\neq 0$, $\zeta$, $\eta$ and $\xi$ are complex valued. Then the index mismatch is neither purely imaginary nor real. The two polarization eigenmodes still experience different phase shifts and different extinctions but the calculations and physical pictures lack the previous simplicity.

\section{The CBS effect}\label{sec_cbsb}

\subsection{Independent scattering approximation}

At low optical density $n\lambda^3 \ll 1$, a semi-classical description of propagation along scattering paths consisting of rays between consecutive scatterers is justified. For resonant scatterers, it implies $k\ell \gg 1$. As a consequence, scattering paths involving different scatterers are uncorrelated and the associated interference averages to zero. Recurrent scattering sequences (visiting a given scatterer more than once) can also be neglected defining the independent scattering approximation (ISA) \cite{AdBart}. In this regime, the wave amplitude A is constructed as the coherent superposition $A =\sum_{\mathcal{P}} A_\mathcal{P}$ of the partial waves scattered along all quasi-classical scattering paths $\mathcal{P}$ joining the positions of the scatterers. Between two successive scatterers the partial waves experiences the effective medium. In the ISA regime, the scattering amplitude $A_\mathcal{P}$ is thus computed using two building blocks, the scattering by an individual atom and the coherent propagation.

\subsection{Multiple scattering and interference}

The average intensity of the wave $I=\langle |\sum_{\mathcal{P}} A_\mathcal{P}|^2\rangle$ breaks into an incoherent contribution $I_{i}= \sum_{\mathcal{P}} \langle|A_\mathcal{P}|^2\rangle$ and a coherent contribution $I_c= 2\textrm{Re}(\sum_{\mathcal{P},\mathcal{P}'} \langle \overline{A_{\mathcal{P}'}} A_\mathcal{P}\rangle)$. The incoherent contribution itself breaks into the sum of a single scattering contribution $I_s$ and a diffuse one, $I_d$, involving scattering paths containing more than $2$ scatterers. All these contributions depend on the polarization of the incoming light and on the detected polarization of the outgoing light.

As is well known, the disorder average does not scramble two-wave interference effects between scattering loops traveled in opposite directions \cite{Houches,Chakra,Rossum}. This is at the core of the CBS effect where interference between amplitudes associated to reverse scattering paths $\mathcal{P}$ and $\widetilde{\mathcal{P}}$ (i. e. paths with the same sequence of scatterers but traveled in opposite order) contribute a constructive interference in a narrow angular cone around the backscattering direction \cite{CBS,AdBart,Rossum}.

\subsection{Backscattered intensity and the CBS contrast}

In this section, we compute the amplitude of multiple scattering paths and the backscattered intensity. We use the contrast of the interferences to determine the degree of coherence of light, which we express in term of a phase coherence length.

Denoting by $\vartheta$ the angle between $\bk$ and the outgoing wave vector $\bk'$, the total average backscattered signal is $I(\vartheta) = I_s(\vartheta)+I_d(\vartheta)+I_c(\vartheta)$ where:
\begin{equation}\label{eq_icoh}
I_c(\vartheta)=2\sum_{\mathcal{P}\geq 2} \textrm{Re} \mv{\overline{a_{\mathcal{P}}}a_{\widetilde{\mathcal{P}}} \, e^{i(\bk+\bk')\cdot\bR_{\mathcal{P}}}}.
\end{equation}
Here $\mathcal{P}\geq2$ means that scattering paths with at least two scatterers are included in the sum, $\bR_p$ being the vector joining the endpoints of path $\mathcal{P}$. The CBS signal $I_c$ varies on a very small angular scale $\sim1/(k\ell)\ll 1$, whereas the angular variations of $I_s$ and $I_d$ follows the Lambert's law and takes place over an angular range of order 1 radian. The incoherent contributions appear to be constant at the angular scale of the CBS cone and can be safely evaluated at $\vartheta=0$.

As a two-wave interference, the CBS signal gives access to the degree of coherence of the outgoing wave and, in turn, at the coherence length of the scattering medium. The interference contrast is quantified by the CBS enhancement factor $\alpha=1+I_c/(I_s+I_d)$ computed at $\vartheta = 0$. which is the ratio of the total intensity at exact backscattering to the total intensity out of the backscattering cone. As single scattering events do not participate to the interference process, they decrease the contrast even if no dephasing mechanism is at work. When $I_s$ can be made to vanish or negligible, the coherence loss is directly associated to the ratio $I_c/I_d$. For classical scatterers, $I_s=0$ in the helicity-preserving polarization channel and reciprocity arguments show that $I_c=I_d$ and $\alpha=2$ in the absence of a magnetic field \cite{StructInt,AdBart,StructInt2}. When a magnetic field is present, $I_c < I_d$ and the coherence length of the medium becomes finite \cite{CBSBexp,CBSBthMartinez,CBSBthBart}. The situation for atoms with nonzero spin in the ground state proves more subtle and will be addressed in the next Section.

\subsection{The intricacies of scattering under a magnetic field}

Consider a scattering path $\mathcal{P}$ containing $s$ atoms, located at ${\br}_i$, whose initial and final magnetic numbers are $m_i$ and $m_i'$ ($i=1,..., s$). The incoming light angular frequency is $\omega$ and its polarization vector $\bs\epsilon$. Along path $\mathcal{P}$, light propagates to the first atom, is scattered, propagates to the second atom and so on. After the $s$-th scattering event, light exits the medium and is detected in the polarization channel $\bs\epsilon'$. One {\it crucial} point is that the internal Zeeman state of an atom with a degenerate ground state can change under scattering, which means, when a magnetic field is present, that the scattered photon can have a different frequency that the incoming one.  In other words, single scattering under a magnetic field is {\it inelastic}, the frequency change being of the order of $\mu B$. When $\mu B$ is larger than or comparable to $\delta$ and/or $\Gamma$, this effect is not negligible: the effective medium and the magneto-optical effects felt by the photon depend on its frequency.
 
If $\omega'$ denotes the outgoing frequency, the amplitude associated to path $\mathcal{P}$ reads:
\begin{eqnarray}\label{eq_adir}
A_\mathcal{P} &=& \overline{\bs\epsilon'}\cdot G_{\omega'}(\br'-\br_{s})t_{m_{s}'m_{s}}G_{\omega_{s-1,s}}(\br_{s}-\br_{s-1})\nonumber\\
&&\ldots  G_{\omega_{1,2}}(\br_{2}-\br_{1})t_{m_{1}'m_{1}}G_\omega(\br_{1}-\br)\cdot\bm\epsilon
\end{eqnarray}
where 
\begin{equation}
\omega_{i,i+1}=\omega+g\mu B\sum_{a=1}^{i}(m_{a}-m_{a}')
\end{equation}
is the frequency of light between the $i$-th and the $(i+1)$-th scatterers. The tensor $t_{m'm}$ has been defined in (\ref{eq_t}).

For the reverse path $\widetilde{\mathcal{P}}$, the incoming angular frequency is still $\omega$ and, by energy conservation, the outgoing angular frequency is still $\omega'$. The amplitude associated to $\widetilde{\mathcal{P}}$ reads:
\begin{eqnarray}\label{eq_arev}
A_{\widetilde{\mathcal{P}}} &=& \overline{\bs\epsilon'}\cdot G_{\omega'}(\br-\br_{1})t_{m_{1}'m_{1}}G_{\omega_{2,1}}(\br_{1}-\br_{2})\ldots \nonumber\\
&& G_{\omega_{s,s-1}}(\br_{s-1}-\br_{s})t_{m_{s}'m_{s}}G_\omega(\br_{s}-\br')\cdot\bm\epsilon 
\end{eqnarray}
with
\begin{equation}
\omega_{i+1,i}=\omega+g\mu_{B}B\sum_{a=i+1}^{p}(m_{a}-m_{a}')
\end{equation}

The angular frequency of light traveling between atoms $i$ and ($i+1$) is $\omega_{i,i+1}$ for path $\mathcal{P}$ and $\omega_{i+1,i}$ for path $\widetilde{\mathcal{P}}$. In general, they {\it differ} and satisfy $\omega_{i,i+1}+\omega_{i+1,i}=\omega+\omega'$. As the magnetic field introduces an explicit difference between $A_\mathcal{P}$ and $A_{\widetilde{\mathcal{P}}}$, the interference contrast will be reduced unless the change of frequency does not occur or is unlikely. This happens for example when the atom does not change its internal state under scattering ($m_i=m_i'$). The conditions for this situation will be examined in the next Section.

As a consequence of this frequency change, the average over the internal degrees of freedom involves the whole scattering path. Indeed, because of the magneto-optical effects and the frequency change under scattering, the atomic internal and external degrees of freedom are intricated in a complicated way: after a scattering event takes place, the location of the next one depends on the value of the scattering mean free path, hence on the frequency of the emitted photon, hence on the change or not of internal state. In such a situation, $I_s$, $I_d$ and $I_c$ can only be computed numerically. In practice one calculates the dimensionless bistatic coefficients $\gamma_x = 4\pi\mathcal V^2\omega^2I_x/(4\pi^2\mathcal{A})$ ($x=s,d,c$), where $\mathcal{A}$ is the illuminated area. 

\subsection{Monte-Carlo simulation}\label{susec_mc}

Analytical results about the properties of the CBS cone can be obtained only in specific cases. In the absence of a magnetic field, the problem has been exactly solved for vector waves in a random medium of Rayleigh scatterers with a uniform density and a slab geometry \cite{Ozrin}. Following the same lines, the solution has been extended to quasi-resonant atomic scatterers with degenerate ground states \cite{Benoit}. For other geometries, numerical calculations are necessary \cite{mc}.

In the presence of a magnetic field, reference \cite{CBSBthBart} contains a generalization of the analytical methods developed in \cite{Ozrin} for Rayleigh scatterers but unfortunately fails to describe some aspects of the experimental results reported in \cite{CBSBexp}. This has to be related to the approximations done to compute the CBS cone, and {\it in fine} to the complexity of the exact calculation. For atomic scatterers, the intrication between the external and internal degrees of freedom makes it almost impossible.

The average intensity $\langle|A^{(s)}_{\mathcal{P}}|^2\rangle$ contributed by a scattering path $\mathcal{P}$ with $s$ scatterers contains an average over the positions of the scatterers, i.e. a $3s$-uple integral, and an average over the internal degrees of freedom, i.e. a sum over the final and initial Zeeman sub-levels of each scatterer in the ground state. To compute this multiple integral and these sums, we use a Monte-Carlo simulation able to extract at the same time $A^{(s)}_\mathcal{P}$ and $A^{(s)}_{\widetilde{\mathcal{P}}}$ for a large number of paths. This numerical simulation allows us to take into account some experimental constraints, such as the shape and the density profile of the atomic cloud or the finite spectral width of the laser probe. Its  principle is as follows:\\

\noindent {\bf Step 1.} A photon with frequency $\omega$, wave vector $\bk=k \hat{{\bf u}}$ and polarization $\bs\epsilon$ enters the atomic cloud. It propagates on a distance $r$ along $\hat{{\bf u}}$ chosen according to the probability distribution 
\begin{equation}\label{eq_P}
P(r)=\frac{1}{\ell(r)}\exp(-\int^{r}\frac{dr'}{\ell(r')}).
\end{equation}
Here $\ell^{-1}(r)=n(r)\sigma(\phi)$ is the inverse local scattering mean free path. It depends on the position of the photon if the atomic number density $n$ is non-uniform. The propagator is the Green's function in the real space (\ref{eq_Gr}).\\
\noindent {\bf Step 2.} The photon at position $r\hat{{\bf u}}$ from the entrance point in the cloud is scattered by an atom. The atomic initial and final states $\ket{Fm}$ and $\ket{Fm'}$ are chosen randomly with a uniform probability distribution. The scattering operator is $t_{m'm}$, eq.(\ref{eq_t}). It transforms the incident polarization into the scattered polarization.\\
\noindent {\bf Step 3.} The photon angular frequency is changed by $g\mu B(m-m')$.\\
\noindent {\bf Step 4.} A contribution to the single scattering bistatic coefficient $\gamma_s$ is computed: the scattered photon is propagated along the backscattering direction $\vartheta$ until it exits the atomic gas yielding the amplitude from which a contribution to $\gamma_s$ is obtained.\\
\noindent {\bf Step 5.} The scattered photon is propagated towards a second scatterer. The direction of propagation is chosen according to an isotropic probability distribution to save computation time. The propagation distance is computed with the help of the distribution law eq.(\ref{eq_P}), the change of frequency and polarization being taken into account.\\
\noindent {\bf Step 6.} The doubly scattered photon is propagated along the backscattering direction $\vartheta$ until it exits the atomic gas yielding the amplitude from which a contribution to $\gamma^{(2)}_d$ is obtained.\\
\noindent {\bf Step 7.} A photon, identical to the incident one, enters the atomic cloud, propagates along the previous double scattering path in reverse order, and exits the medium in the backscattering direction $\vartheta$. The scatterers experience exactly the same atomic transitions. This yields the amplitude associated to the reverse previous double scattering path from which, together with Step 6, a contribution to $\gamma^{(2)}_c$ is obtained.\\
\noindent {\bf Step 8.} The process is continued (triple scattering, etc) until the photon finally exits the atomic cloud.\\
\noindent {\bf Step 9.} Another incident photon is sent in the cloud and the whole process is repeated as many times as necessary to obtain a good signal-to-noise ratio. Typically, one needs to launch between 10$^6$ and 10$^9$ photons to obtain well converged values for $\gamma_s$, $\gamma_d=\sum_s \gamma^{(s)}_d$ and $\gamma_c = \sum_s \gamma^{(s)}_c$.\\

Up to the statistical errors, this method is quasi-exact and limited only by computer resources in the limit $k\ell\gg 1$.

When the magnetic field vanishes, the results of the Monte-Carlo simulations reported in \cite{mc} are recovered. At large magnetic fields, the modulus of the amplitudes associated to reverse paths are very sensitive to the scattering parameters of the paths. Any change in a direction of propagation modifies significantly the refractive index of the effective medium. As a consequence, the Monte-Carlo simulation needs to average over more and more fluctuating quantities when the magnetic field increases. The statistical error on the total diffuse intensity can be estimated by its standard deviation. It remains smaller than 1$\%$ for small magnetic fields ($\mu B/\Gamma<1$), and smaller than 5$\%$ for magnetic fields up to ($\mu B/\Gamma\simeq 10$) for the results presented in section \ref{susec_mc_sim}.

\section{Restoration of the CBS contrast}\label{sec_rest}

In the following, we apply the results of the previous Sections to compute the CBS cone for quasi-resonant light propagating in a cold $^{85}$Rb cloud. The frequency of light is chosen close to the frequency of the $F=3\rightarrow F_e=4$ transition of the D2 line (wavelength $\lambda=780$nm, linewidth $\Gamma/(2\pi)=5.9$MHz). The Land\'e factors of the ground and excited states are $g=1/3$ and $g_e=1/2$. For this particular transition, a Zeeman shift $\mu B=\Gamma$ corresponds to $B=4.2$G.

At $B=0$, CBS experiments have reported very low enhancement factors, e. g. $\alpha\approx1.05$ in the helicity preserving channel \cite{LowEnFac1,LowEnFac2}. This is in marked contrast with experiments with spherically-symmetric classical scatterers where reciprocity guarantees $\alpha$ takes its maximal value $2$ in the same polarization channel \cite{Factor2}. A detailed analysis shows that the low $\alpha$ value observed with cold atoms comes from an imbalance between the amplitudes associated to reverse paths \cite{StructInt}. This imbalance is noticeably absent for a $F=0\to F_e=1$ transition where $\alpha=2$ is recovered \cite{Bidel}. It is our goal in this section to show that the interference contrast can be fully restored with the help of an external magnetic field.

\subsection{Filtering out a closed transition}\label{susec_setup}

The key idea to restore the CBS contrast in the helicity-preserving channel is simply to lift the degeneracy of the atomic ground state and to filter out a closed transition. This is done by splitting the Zeeman sub-levels with an external magnetic field (Zeeman effect) and by shining the atomic cloud with a light wave which is resonant with the $\ket{F=3,m=3}\rightarrow\ket{F_e=4,m_e=4}$ transition. To achieve this, one needs to impose $\delta=(4g_e-3g)\mu B = \mu B$. This transition is closed since an atom in the excited state $\ket{44}$ can only make a  transition to the ground state $\ket{33}$. At sufficiently large $B$, the other Zeeman sub-levels of the ground and excited states are sufficiently split away and are out of resonance, meaning that the $\ket{33}\rightarrow\ket{44}$ transition is isolated. Thus, at large $B$, the atomic cloud consists of (i) atoms which are in the sub-state $\ket{33}$ and can scatter light, and (ii) atoms which are not in the sub-state $\ket{33}$ and cannot scatter light because the frequency is too far-detuned from the other transitions. These $\ket{33}$-scatterers behave like effective two-level atoms which can only absorb and emit $\sigma_+$ radiation, i. e. light with positive helicity along $\hat{{\bf B}}$.

Under these circumstances, it makes no difference for light to travel a scattering path in one direction or the other. The multiple scattering amplitudes associated to any path $\mathcal{P}$ and to its reverse partner $\widetilde{\mathcal{P}}$ are equal and the CBS contrast is restored. This restoration is expected to be most spectacular in the helicity-preserving channel, because it is in this channel that the contrast is the lowest without any magnetic field. If the incident light beam is parallel to $\bB$ and $B$ is sufficiently large, it is easy to see that one only gets a non-vanishing CBS signal in the helicity non-preserving polarization channel. We will thus choose in the following the "Cotton-Mouton" configuration where $\bB$ is perpendicular to the incident light beam and analyze the CBS signal in the helicity-preserving polarization channel. 

\subsection{Small magnetic fields}

From the previous discussion, it seems that the contrast restoration only occurs at sufficiently large $B$. In fact, it turns out that the contrast restoration even starts at small magnetic fields and gets larger as $B$ is increased. To demonstrate this, we study analytically the single and double scattering signal in a uniform, semi-infinite medium in the limit $\mu B\ll \Gamma$ (meaning $B\ll2.1$ G for $^{85}$Rb). It is then possible to neglect the magneto-optical effects, and to propagate photons with the propagator (\ref{eq_Gr}) evaluated at $B=0$. This will be justified below in Section \ref{susec_imoe}. Expanding the scattering matrices $t_{m'm}$ at second order in $\phi_B=2\mu B/\Gamma$, and noticing that $\phi\approx\phi_B+i\phi_B^2$ at same order, we then use expressions (\ref{eq_adir}) and (\ref{eq_arev}) to compute the single and double scattering amplitudes amplitudes. In the chosen geometry, the average over the external degrees of freedom when computing $\gamma_s$, $\gamma^{(2)}_d$ and $\gamma_c^{(2)}$ can be done analytically \cite{StructInt2}. The calculations have been made here with the symbolic calculation software \textit{Maple}\texttrademark {} and yields:
\begin{eqnarray}
\gamma_s/\gamma_d^{(2)}&=&0.305+0.468\phi_B^2 \\
\gamma_c^{(2)}/\gamma_d^{(2)}&=&0.217+0.143\phi_B^2 \\
\alpha^{(2)}&=&1.166+0.048\phi_B^2
\end{eqnarray}
 As one can see, both quantities increase with $B$, meaning that the coherence length of the system is increased.

\subsection{Monte-Carlo simulations}\label{susec_mc_sim}

\begin{figure}
\begin{center}
\includegraphics[width=8cm,height=6cm]{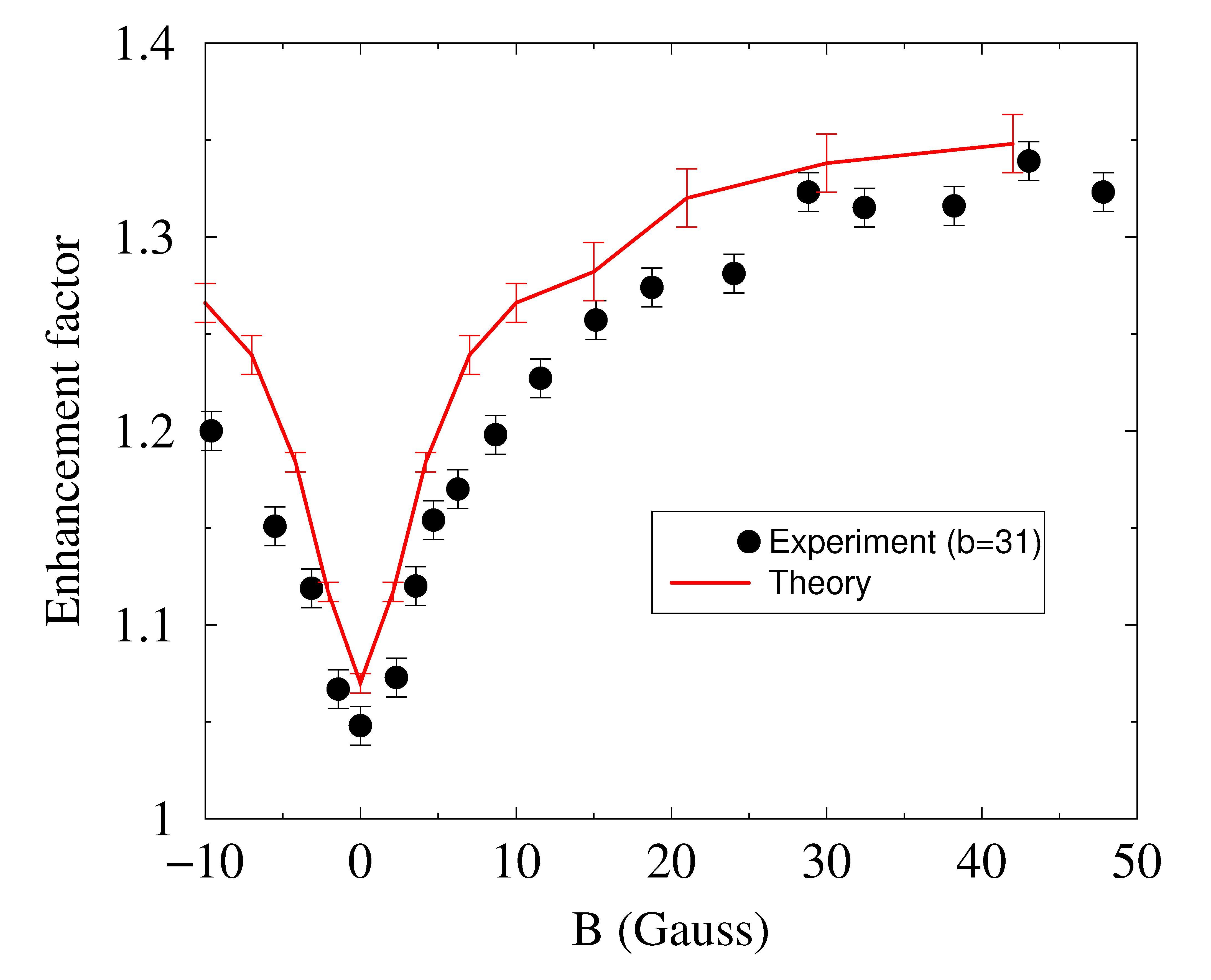}
\end{center}
\caption{\label{fig_alpha}  Plot of the CBS enhancement factor $\alpha$ as measured in the parallel helicity channel $h\,{\parallel}\, h$ for different values of $B$ (circles) for light backscattered by a cold $^{85}$Rb atomic cloud in the Cotton-Mouton configuration $\bk \perp \bB$. For each value of $B$, the light is tuned on resonance with the $\ket{33}\to\ket{44}$ transition ($\delta=\mu B$). The spherically-symmetric atomic cloud is characterized by a gaussian density and an optical thickness $b=31$ when $\delta=0$ and $B=0$. One witnesses a dramatic {\em increase} of the CBS contrast compared to the situtation at $B=0$ despite the fact that the time-reversal symmetry is broken. The solid line is the result of the Monte-Carlo simulation with no adjustable
parameters% other than the laser linewidth and the cloud shape
.}
\end{figure}

When the magnetic field is neither small nor large, there is no simple approximation that allows to compute analytically the bistatic coefficients, but they can at least be computed numerically with the help of the Monte-Carlo simulation described in Section \ref{susec_mc}. It is then possible to take into account a more realistic model of the atomic cloud than a semi-infinite uniform medium. We present results here for a spherically-symmetric atomic cloud with gaussian density and optical thickness $b=31$ (measured at $B=0$ and $\delta=0$). We take a laser probe beam with spectral width equal to $0.3\Gamma$. This allows a realistic comparison with our experiment done at a fixed total number of atoms. The computed values of $\alpha$ are compared to the experimental ones for various values of $B$ in Fig.\ref{fig_alpha}. As one can see, $\alpha$ increases with $B$, starting from $\alpha=1.05$ at $B=0$ up to $\alpha\simeq 1.35$ at $B=40$ G. The agreement between theory (solid line) and experiment (circles) is quite satisfactory. This shows that the Monte-Carlo simulation contains the essential ingredients that play a role in the restoration of the contrast. In the following, we will rest on the results of the Monte-Carlo simulation to elucidate the mechanisms at work by computing quantities which are not accessible to experiment, e. g. the bistatic coefficients for each scattering order.

\begin{figure}
\begin{center}
\includegraphics[width=9cm,angle=0]{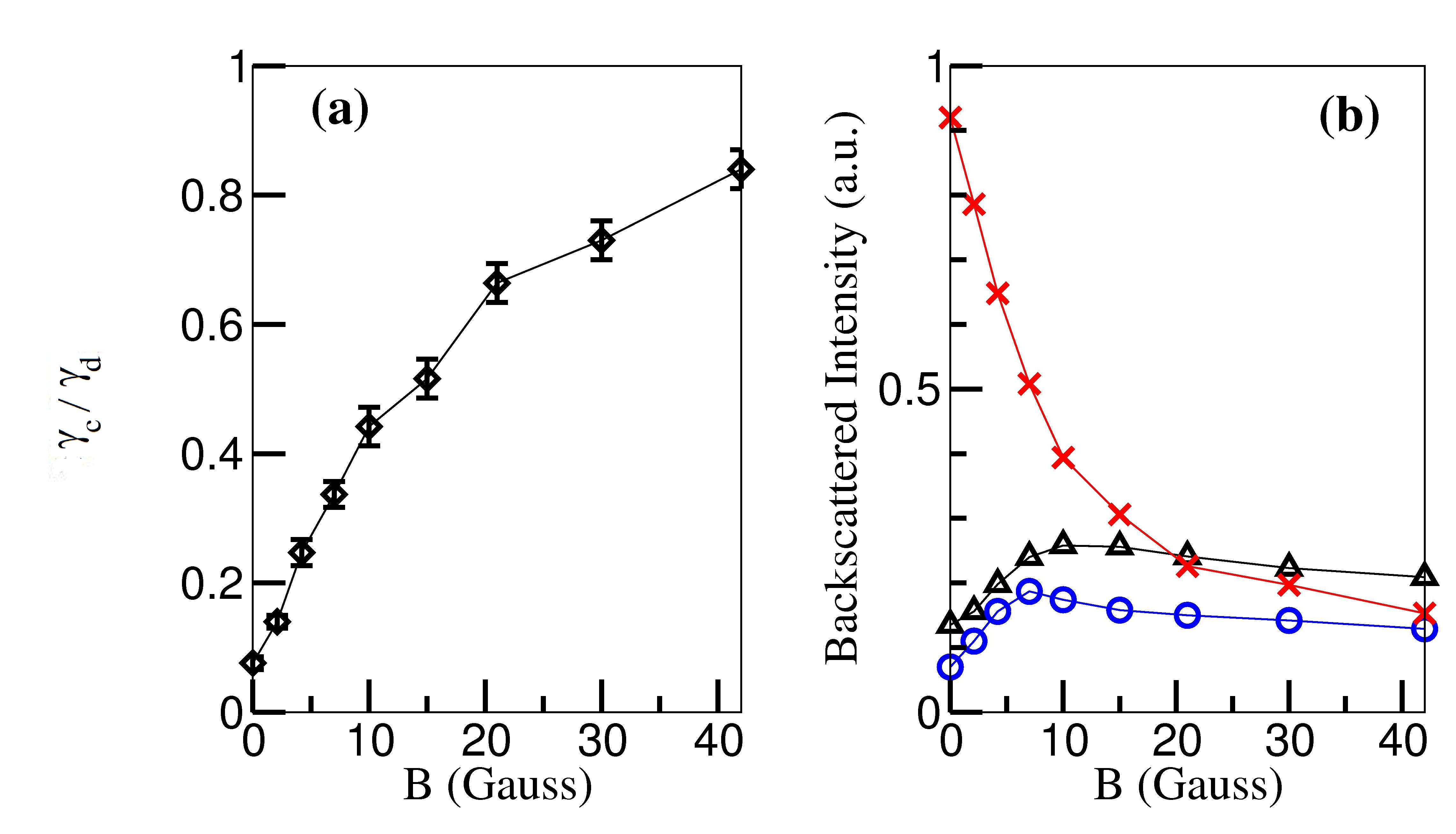}
\end{center}
\caption{\label{fig_int} Results of the Monte-Carlo simulations under the same experimental conditions as Fig.\ref{fig_alpha}. (a) The coherence ratio $\gamma_c/\gamma_d$ increases with $B$. (b) Plots of the bistatc coefficients $\gamma_s$ (triangles), $\gamma_d$ (crosses) and $\gamma_c$ (circles) in the backward direction as a function of $B$. The solid lines are drawn to guide the eyes. As one can see, $\gamma_c$ first increases than decreases with $B$ without varying too much. At the same time, $\gamma_d$ decreases strongly. At large $B$, $\gamma_c=\gamma_d$ and the coherence is restored. However the enhancement factor $\alpha <2$ because $\gamma_s$ is not negligible.}
\end{figure}

In Fig.\ref{fig_int}a, we plot $\gamma_c/\gamma_d$ as a function of $B$. This ratio is a measure of the degree of coherence of the outgoing light. This ratio grows when $B$ increases, and tends to 1 for large $B$ (not shown in the figure). This confirms that the contrast of the interference, and in turn the coherence of the outgoing light, is actually restored by a magnetic field.

In Fig.\ref{fig_int}b we show how $\gamma_s$ (triangles), $\gamma_d$ (crosses) and $\gamma_c$ (circles) change with $B$. $\gamma_d$ strongly decreases strongly with $B$ because the atomic scattering cross section itself decreases. Meanwhile, $\gamma_c$ increases for magnetic fields up to 8 G. This shows the efficiency of the mechanism restoring interference. At larger fields however, the decrease of the scattering cross section takes over and $\gamma_c$ decreases again, although slower than $\gamma_d$. At large $B$, $\gamma_c$ and $\gamma_d$ tends to the same value (yielding a perfect coherence) but are outgrown by $\gamma_s$, meaning $\alpha<2$. This behavior is not generic: at larger optical thickness, $\gamma_s$ would have been smaller than $\gamma_d$.

We also mention that the values of the bistatic coefficients are independant of the value of $k\ell\gg 1$. However, the angular width of the backscattering cone depends on $k\ell$.

\subsection{Impact of Faraday and Cotton-Mouton effects}\label{susec_imoe}

\begin{figure}
\begin{center}
\includegraphics[width=7cm]{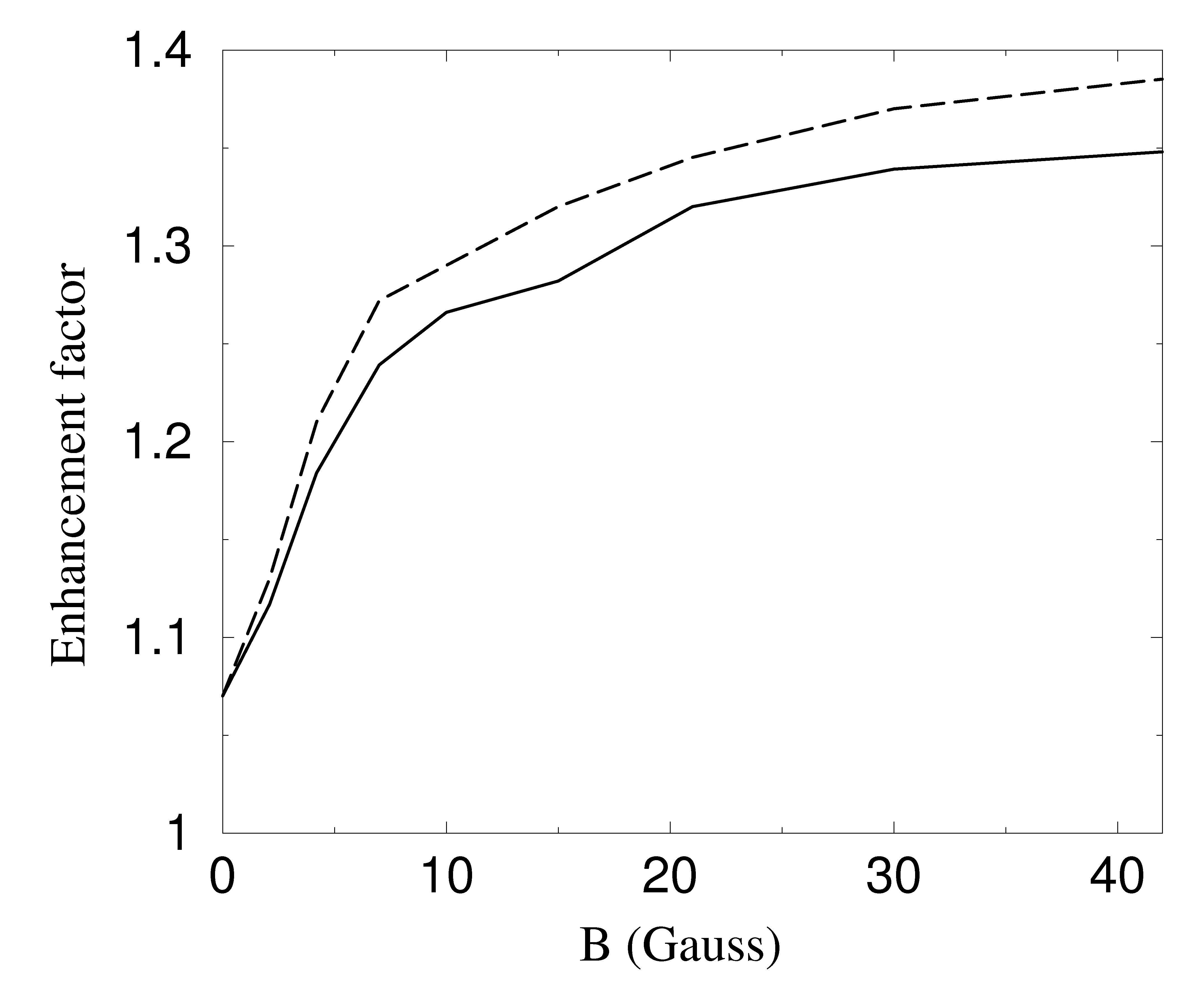}
\end{center}
\caption{\label{fig_moe} Impact of magneto-optical effects occurring during propagation (Faraday and Cotton-Mouton effects) on the CBS enhancement factor $\alpha$. The solid line is the theoretical curve obtained in Fig.\ref{fig_alpha}. The dotted line is the enhancement factor $\tilde{\alpha}$ calculated by discarding the magneto-optical effects. as one can see, the Faraday and Cotton-Mouton effects do decrease the contrast but their detrimental effect is counter-balanced and beaten by an efficient mechanism restoring the contrast. This mechanism is the modification of the scattering properties of the atoms which, because of the Zeeman splitting, behave more and more as effective two-level atoms when $B$ is increased and the light is tuned on resonance with a closed transition.}
\end{figure}

To study the impact of magneto-optical effects on the enhancement factor, the simplest way is to discard them in the Monte-Carlo simulation, and compare the obtained result with the experimental data in Fig.\ref{fig_alpha}. The magneto-optical effects distort the polarization of a propagating wave. Discarding them means here that a polarization is propagated without distorsion, only with attenuation. To ensure energy conservation, the attenuation length (i.e. the scattering mean free path) must be equal to $\ell=1/(n(\br)\sigma(\phi))$ where $n(\br)$ is the local density of atoms and $\sigma(\phi)$ the total scattering cross section of an atom given by equation (\ref{eq_sigma}). All other parameters in the simulation are left unchanged.

Fig.\ref{fig_moe} shows the plot of this expurgated enhancement factor $\tilde{\alpha}$ as a function of $B$ (dotted line), together with its quantitative comparison to the real theoretical curve borrowed from Fig.\ref{fig_alpha} (solid line). For $B\lesssim4$G ($\mu B\lesssim\Gamma$), the impact of Faraday and Cotton-Mouton effects is negligible. For larger fields, the true value $\alpha$ is slightly lower than $\tilde{\alpha}$, the two curves being roughly parallel to each other. This means that magneto-optical effects do decrease the phase coherence of the sample but this detrimental effect is counter-balanced and beaten by a more powerful mechanism restoring coherence. As a matter of fact, at large $B$ and after the first scattering event, a single eigenmode can propagate in the scattering medium, exemplifying why phase or extinction differences between the polarization eigenmodes cannot scramble the contrast. Thus, the phenomenon explaining the restoration of the CBS contrast with $B$ is really the modification of the scattering properties of the atoms which, because of the Zeeman splitting, behave more and more as effective two-level atoms when $B$ is increased and the light is tuned on resonance with a closed transition. This is in marked contrast with classical scatterers where no such mechanism counter-balancing the detrimental magneto-optical effects does exist.

\subsection{Influence of higher and higher scattering orders}

\begin{figure}
\begin{center}
\includegraphics[width=7cm]{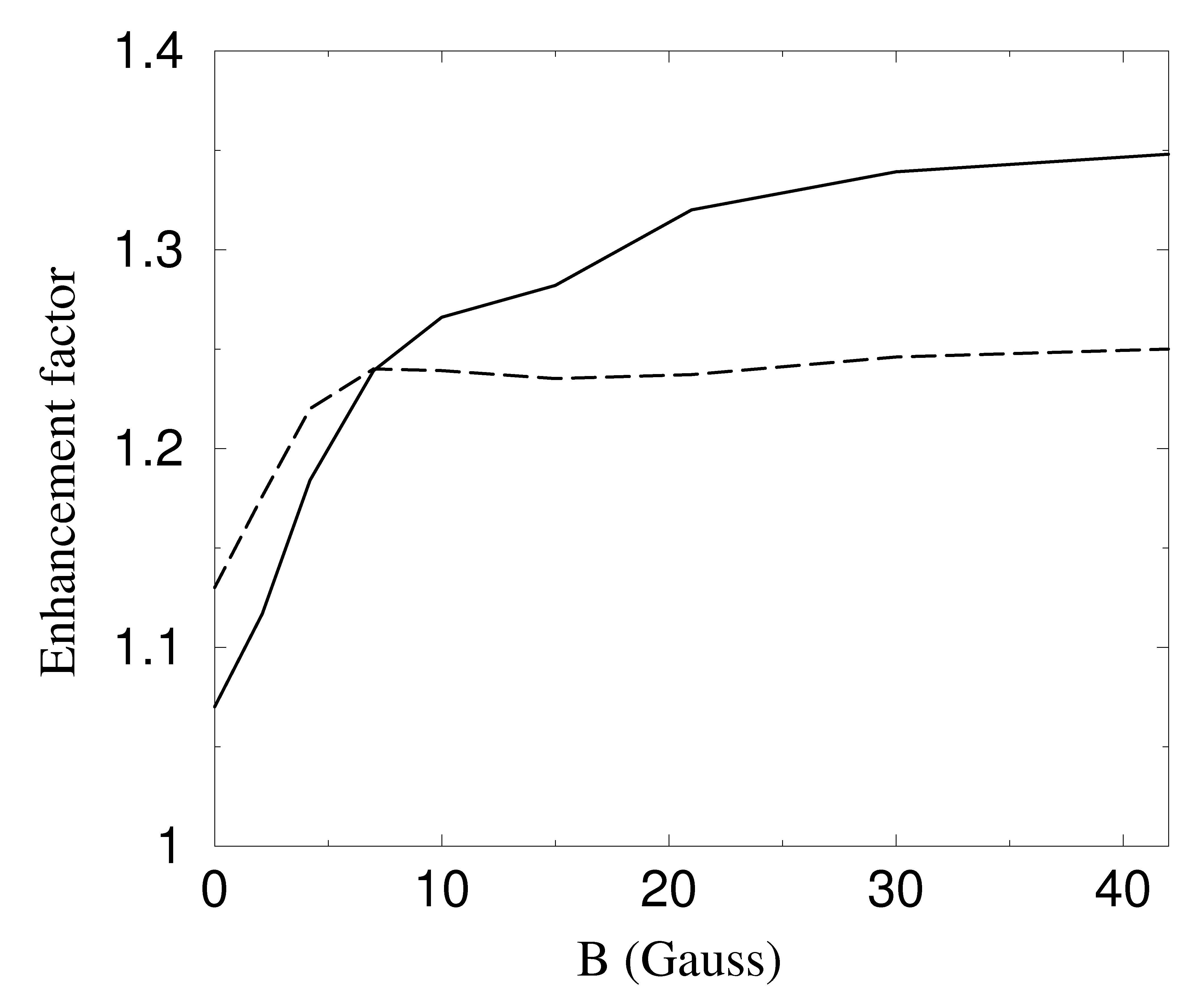}
\end{center}
\caption{\label{fig_lso} Influence of higher and higher scattering orders on the CBS enhancement factor. The incoming light and atomic cloud parameters are given in Fig.\ref{fig_alpha}. We compare the theoretical enhancement factor borrowed from Fig.\ref{fig_alpha}) with the one obtained by only considering single and double scattering orders. At small $B$, high scattering orders contribute mostly to the diffuse signal. At large $B$, they contribute equally to the coherent and diffuse signals, making the CBS cone height increase.}
\end{figure}

When $B=0$, the CBS effect observed with atoms having a degenerate ground state is dominated by double scattering paths, while higher-order scattering paths contribute significantly to the diffuse background \cite{LowEnFac2}. This is shown in the first line of Table \ref{tab1}. As $B$ increases, the contrast is restored and higher and higher scattering orders contribute significantly both to $\gamma_c$ and $\gamma_d$, see second line of Table \ref{tab1}.
\begin{table}[h]
\begin{center}
\begin{tabular}{|lr|cccc|}\hline \noalign{\smallskip}
Scattering order&&2&3&4&5\\ \hline
$\gamma_c/\gamma_d$&B=0G&0.21&0.08&0.04&0.02\\ %\cline{2-6}
&B=30G&0.81&0.75&0.69&0.38\\ \hline \noalign{\smallskip}
\end{tabular}
\end{center}
\caption{\label{tab1}  CBS coherence factor $\gamma_c/\gamma_d$ for the first scattering orders when $B=0$G and $B=30$G. The incoming light and atomic cloud parameters are given in Fig.\ref{fig_alpha}. The atomic cloud is spherically symmetric with gaussian density.}
\end{table}
However, at the same time, the optical thickness of the atomic cloud decreases. Higher-order scattering paths become less and less probable and the two effects compete. The dotted line in figure \ref{fig_lso} shows the enhancement factor calculated from the single and double scattering contributions alone. This approximation overestimates the height of the backscattering cone at small magnetic fields, but underestimates it at large magnetic fields. This shows that high scattering orders do  contribute to the CBS cone and cannot be discarded for a quantitative comparison.

\subsection{Role of optical pumping}

\begin{figure}
\begin{center}
\includegraphics[width=7cm]{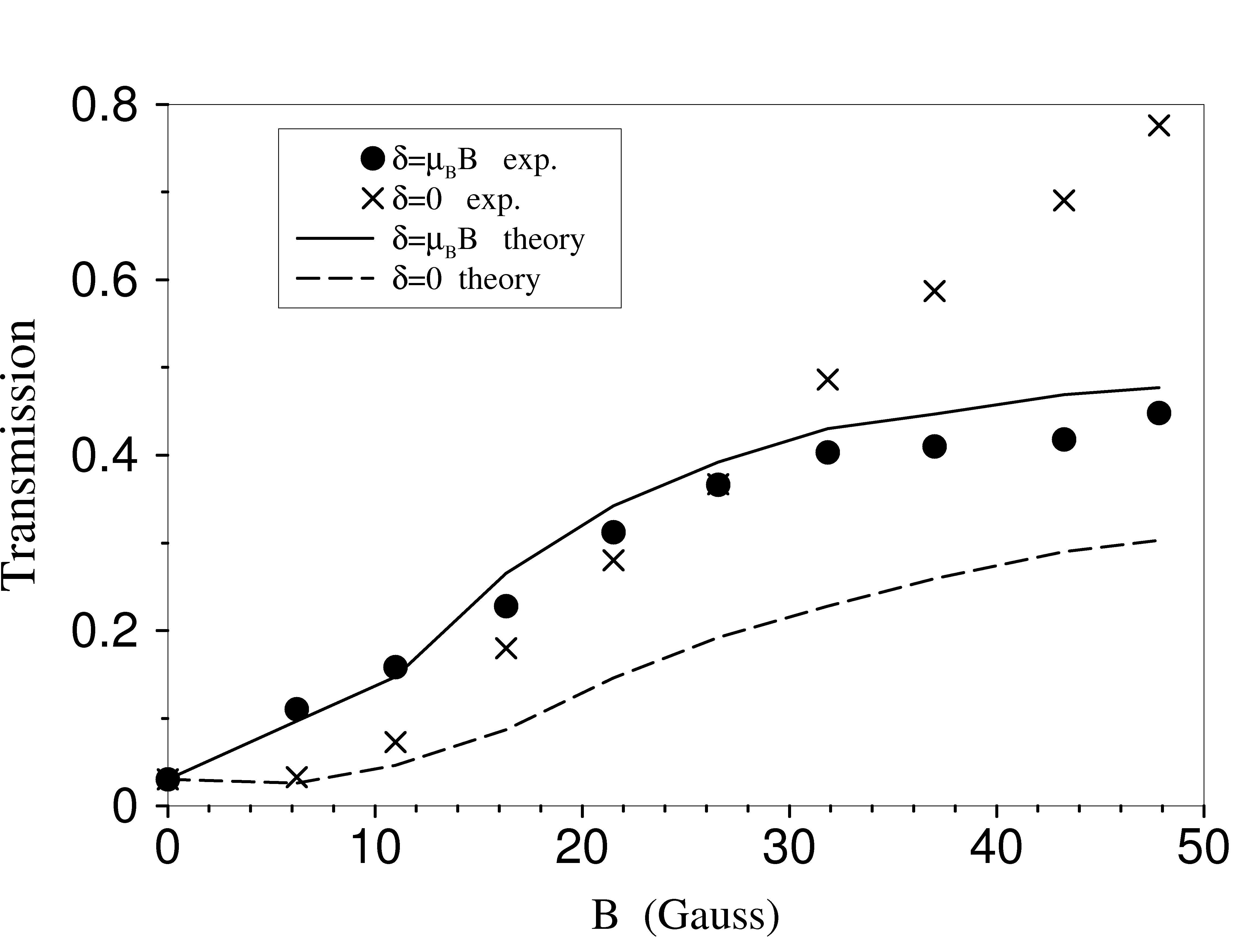}
\end{center}
\caption{\label{fig_trans} Coherent transmission of the atomic cloud as a function of $B$ for an initial optical thickness $b=31$ measured at $B=0$ and $\delta=0$. Crosses: experiment for $\delta=0$. Dashed line: theory for $\delta=0$. Solid circles: experiment for $\delta=\mu B$. Solid line: theory for $\delta=\mu B$. When $\delta=\mu B$ the incoming light is kept on resonance with the closed atomic transition $\ket{33}\rightarrow\ket{44}$. When $\delta=0$, the coherent transmission increases in time and converges to the plotted value. In this case, optical pumping is at work, a situation not accounted for by our theory. For $\delta=\mu B$, no time variation of the coherent transmission is observed. This is a strong indication that optical pumping is negligible in this case.}
\end{figure}

In section \ref{susec_po}, we mentioned that our theory does not take optical pumping into account (though it could be extended to do so). In the present section, we give conclusive evidence that optical pumping is indeed negligible in our experiment by measuring the coherent transmission of the atomic cloud. The results are presented in Fig.\ref{fig_trans}, for an incoming wavevector perpendicular to $\bB$ and a circular incoming polarization. If $\delta=0$, the coherent transmission varies in time and its stationary value is shown in Fig.\ref{fig_trans} as a function of $B$ (crosses). Our Monte Carlo simulation (dotted line) is unable to reproduce these results for $B>10$ G, indicating that optical pumping is indeed present in our sample when $\delta=0$. However, when the incident light beam is kept at resonance with the atomic transition $\ket{33}\rightarrow\ket{44}$ (i.e. $\delta=\mu B$), no time evolution of the coherent transmission is observed. This shows that the populations of the various Zeeman substates are almost constant. The experimental data (circles) are well reproduced by the Monte-Carlo simulation (solid line). This is a strong indication that optical pumping is indeed negligible when the incoming light is continuously kept at resonance with the closed atomic transition. 

\subsection{Coherence length}

\begin{figure}
\begin{center}
\includegraphics[width=7cm,angle=0]{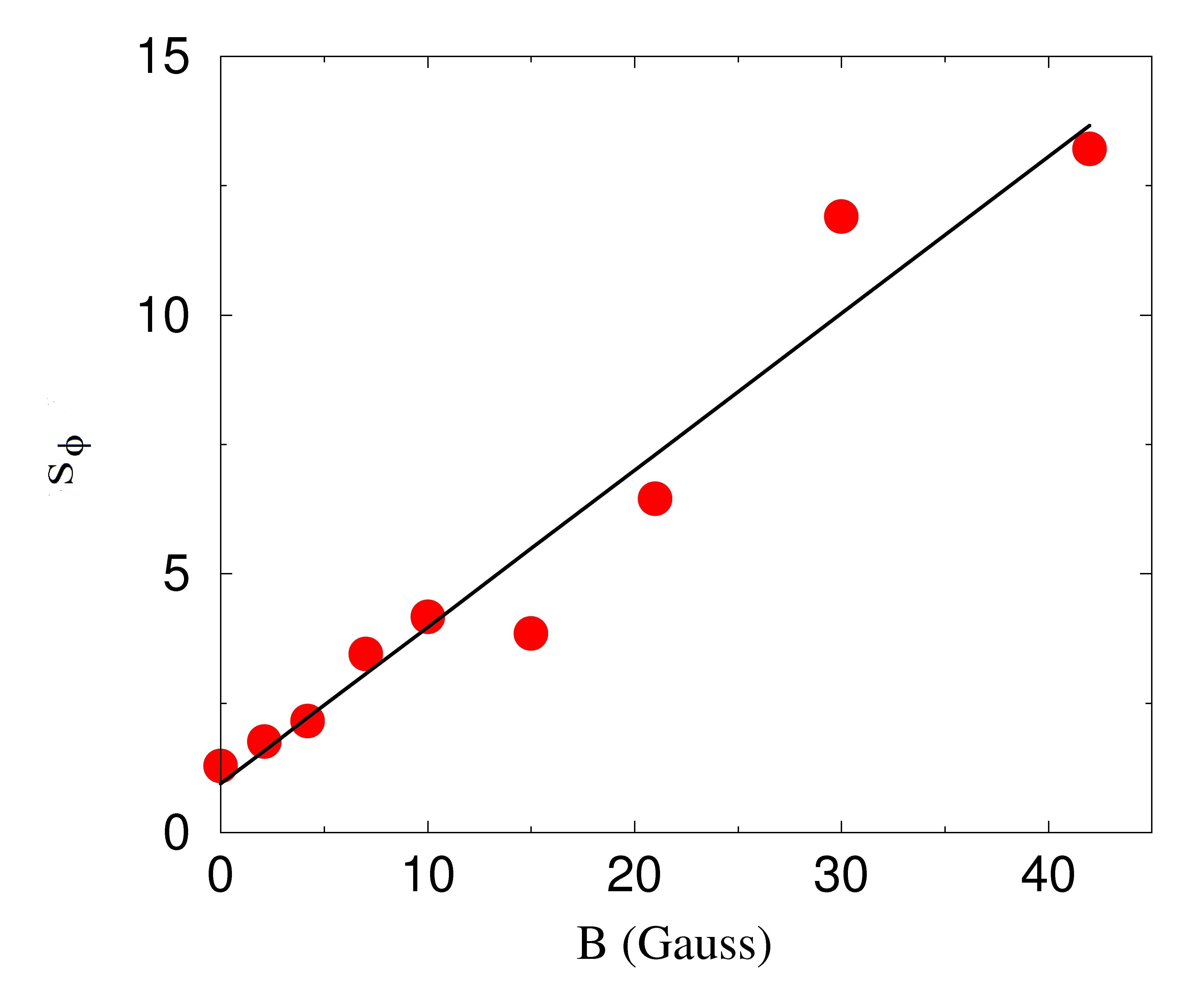}
\end{center}
\caption{\label{fig_Lcoh} Plot of $s_\phi = L_{\phi}/\ell$ as a function of $B$, as extracted from our Monte-Carlo calculation. The incoming light and atomic cloud parameters are given in Fig.\ref{fig_alpha}. The points dispersion reflects the numerical accuracy. As one can see, $s_\phi$ {\emph increases} with $B$, roughly linearly (the
solid line is drawn to guide the eye). This is due to the lifting of the Zeeman degeneracy which make atoms behave like effective two-level atoms when the light is tuned on resonance with a closed transition. This behavior is in sharp contrast with classical scatterers where $s_\phi$ decreases when $B$ increases.}
\end{figure}

The notion of phase coherence length $L_\phi$ is a very important concept in mesoscopic physics. It is the length scale at which, because of some dephasing mechanisms, the interference effects as produced by the medium are effective. The larger is $L_\phi$, the stronger is the impact of interference, and, in the case of the CBS effect, the larger is the CBS contrast. In the case of cold atoms, at $B=0$, the degeneracy of the atomic ground state causes a loss of phase coherence between reversed scattering paths giving rise to a finite value of $L_\phi$ of the order of few mean free paths $\ell$ \cite{StructInt}. The increase of the CBS contrast when a magnetic field is applied is accordingly accompanied by a growth of $L_\phi$. The Monte-Carlo simulation allows us to estimate the phase coherence length in the following way. In the presence of dephasing, the interference term $\gamma_c$ associated to two reversed scattering paths of length $L$ is related to the diffuse term $\gamma_d$ by:
\begin{equation}
\gamma_c \simeq \gamma_d \, e^{-L/L_\phi} = \gamma_d \, e^{-s/s_\phi},
\end{equation}
where $L/\ell=s$, $s$ being the scattering order, and $L_\phi/\ell = s_\phi$.
In Fig.\ref{fig_Lcoh} we plot $s_\phi$ as a function of $B$ as obtained numerically. It increases roughly linearly.

One should note that our definition of the coherence length differs from the usual one where the distance travelled diffusively by the light inside the disordered sample is introduced \cite{Houches}. With this convention, $L\propto\sqrt{s}$ and $L_\phi\propto\sqrt{s_\phi}$.

\subsection{Analogy with paramagnetic impurities in solid-state physics}

The surprising fact that a magnetic field can restore weak localization effects under well chosen circumstances although it breaks time-reversal invariance is already known in solid-state physics \cite{Washburn,Pothier}. In this context, one considers the propagation of electrons inside a metal at low (but finite) temperature, containing paramagnetic impurities. Thermal fluctuations make the spin of these impurities fluctuate in time. The scattering of an electron by such a fluctuating impurity randomizes the electron spin and the weak localization correction to electronic transport are reduced. This is similar to the loss of contrast due to the degeneracy of the atomic ground state. When a large enough magnetic field is applied to the metal, the spins of the impurities are all aligned along $\bB$. The fluctuations of the spin component of the electrons are suppressed and the weak localization corrections to transport are restored.

Finally, in both cases, the magnetic field freezes the internal degrees of freedom and restores interference effect. The main difference with our case is that in solid-state physics, the magnetic field populates a unique spin state, whereas the Zeeman substates of the atoms are equally populated. It could be possible to realize an atomic cloud almost containing only atoms in the ground state $\ket{33}$, by using optical pumping. However, this would only restore the interference between reverse double scattering paths \cite{PolarAt}. Nevertheless, it would be possible to first populate the $\ket{33}$ state, and then to apply an external magnetic field. This would increase the number of atoms participating to the scattering of light. Multiple scattering would then play a more important role and the enhancement factor would be larger than reported in the present article.

\section{Conclusion}

To summarize, we have accurately described in this paper the propagation of light in cold atomic gases in the multiple scattering regime where $k\ell \gg 1$ and in the presence of an applied external magnetic field. In this regime, a semi-classical description is well suited and transport is described through a series of individual scattering events separated by coherent propagation in an effective medium. We have studied the impact of the magnetic field on the scattering of light by atoms with a degenerate ground state (differential and total cross sections) and the magneto-optical effects (Faraday and Cotton-Mouton effects) embodied in the refractive index tensor of the effective medium. Our results generalize previous works \cite{Hanle,VoigtCold}. We have then applied our theory to the study of the coherent backscattering effect and we have shown that the magnetic field can lead to a full restoration of the two-wave interference contrast provided the incoming light is continuously set on resonance with a closed atomic transition as $B$ is increased. The reason for the restoration of contrast is that the atoms behave as effective two-level atoms for which scattering amplitudes associated to reverse scattering paths have the same strength.

\acknowledgements
OS wishes to thank the Centre for Quantum Technologies (CQT) for its kind hospitality. ChM acknowledges
support from the CNRS-CQT LIA FSQL. The
CQT is a Research Centre of Excellence funded by the Ministry
of Education and the National Research Foundation of
Singapore.

%\begin{figure}
%\resizebox{0.75\columnwidth}{!}{%
%  \includegraphics{leer.eps}
%\vspace{5cm}       % Give the correct figure height in cm
%\end{figure}
%\begin{figure*}
%\vspace*{5cm}       % Give the correct figure height in cm
%\end{figure*}

%\begin{table}
%\begin{tabular}{lll}
%\hline\noalign{\smallskip}
%\end{tabular}
% Or use
%\vspace*{5cm}  % with the correct table height
%\end{table}
%
% BibTeX users please use
% \bibliographystyle{}
% \bibliography{}

\end{document}